\documentclass[twocolumn,tighten]{aastex62} 

\usepackage{graphicx}
\usepackage{color}
\DeclareGraphicsExtensions{.eps,.png,.jpg}
\usepackage{epsfig, psfig}

\def\asec{$^{\prime\prime}$}

\def\kms{km s$^{-1}$}

\def\lax{{$\mathrel{\hbox{\rlap{\hbox{\lower4pt\hbox{$\sim$}}}\hbox{$<$}}}$}}
\def\gax{{$\mathrel{\hbox{\rlap{\hbox{\lower4pt\hbox{$\sim$}}}\hbox{$>$}}}$}}
\def\simlt{\lower.5ex\hbox{$\; \buildrel < \over \sim \;$}}
\def\simgt{\lower.5ex\hbox{$\; \buildrel > \over \sim \;$}}

\def\sb{mag~arcsec$^{-2}$}
\def\solum{$L_\odot$}
\def\solmass{$M_\odot$}

\journalinfo{The Astrophysical Journal, 859:37}
\reportnum{\small{https://doi.org/10.3847/1538-4357/aabbae}}
\received{2017 September 19}
\revised{2018 March 29}
\accepted{2018 March 31}
\published{2018 May 21}

\begin{document}
\title{\large\bf{Low Metallicities and Old Ages for Three Ultra-Diffuse Galaxies in the Coma Cluster}}

\author[0000-0002-4267-9344]{Meng Gu}
\affiliation{Department of Astronomy, Harvard University, Cambridge, MA 02138, USA}
\author[0000-0002-1590-8551]{Charlie Conroy}
\affiliation{Department of Astronomy, Harvard University, Cambridge, MA 02138, USA}
\author[0000-0002-9402-186X]{David Law}
\affiliation{Space Telescope Science Institute, 3700 San Martin Drive, Baltimore, MD 21218, USA}
\author[0000-0002-8282-9888]{Pieter van Dokkum}
\affiliation{Astronomy Department, Yale University, New Haven, CT 06511, USA}
\author[0000-0003-1025-1711]{Renbin Yan}
\affiliation{Department of Physics and Astronomy, University of Kentucky, 505 Rose Street, Lexington, KY 40506-0057, USA}
\author[0000-0002-6047-1010]{David Wake}
\affiliation{Department of Physical Sciences, The Open University, Milton Keynes, MK7 6AA, UK}
\author[0000-0001-9742-3138]{Kevin Bundy}
\affiliation{Department of Astronomy and Astrophysics, University of California, Santa Cruz, CA 95064, USA}
\author[0000-0001-9467-7298]{Allison Merritt}
\affiliation{Astronomy Department, Yale University, New Haven, CT 06511, USA}
\author[0000-0002-4542-921X]{Roberto Abraham}
\affiliation{Department of Astronomy and Astrophysics, University of Toronto, 50 St. George Street, Toronto, ON M5S 3H4, Canada}
\author[0000-0001-5310-4186]{Jielai Zhang}
\affiliation{Department of Astronomy and Astrophysics, University of Toronto, 50 St. George Street, Toronto, ON M5S 3H4, Canada}
\affiliation{Canadian Institute for Theoretical Astrophysics, University of Toronto, 60 St. George Street, 14th floor, Toronto, ON M5S 3H8, Canada}
\affiliation{Dunlap Institute for Astronomy and Astrophysics, University of Toronto, 50 St. George Street, Toronto, ON M5S 3H4, Canada}
\author[0000-0002-3131-4374]{Matthew Bershady}
\affiliation{Department of Astronomy, University of Wisconsin-Madison, 475N. Charter Street, Madison WI 53703, USA}
\author[0000-0002-3601-133X]{Dmitry Bizyaev}
\affiliation{Apache Point Observatory, P.O. Box 59, Sunspot, NM 88349, USA}
\affiliation{Sternberg Astronomical Institute, Moscow State University, Moscow, Russia}
\author[0000-0001-6128-659X]{Jonathan Brinkmann}
\affiliation{Apache Point Observatory, P.O. Box 59, Sunspot, NM 88349, USA}
\author[0000-0002-7339-3170]{Niv Drory}
\affiliation{McDonald Observatory, The University of Texas at Austin, 1 University Station, Austin, TX 78712, USA}
\author{Kathleen Grabowski}
\affiliation{Apache Point Observatory, P.O. Box 59, Sunspot, NM 88349, USA}
\author[0000-0003-0846-9578]{Karen Masters}
\affiliation{Institute of Cosmology \& Gravitation, University of Portsmouth, Dennis Sciama Building, Portsmouth, PO1 3FX, UK}
\author[0000-0002-2835-2556]{Kaike Pan}
\affiliation{Apache Point Observatory, P.O. Box 59, Sunspot, NM 88349, USA}
\author{John Parejko}
\affiliation{Department of Astronomy, University of Washington, Box 351580, Seattle, WA 98195, USA}
\author[0000-0002-5908-6852]{Anne-Marie Weijmans}
\affiliation{School of Physics and Astronomy, University of St. Andrews, North Haugh, St. Andrews KY16 9SS, UK}
\author[0000-0002-9808-3646]{Kai Zhang}
\affiliation{Lawrence Berkeley National Laboratory, 1 Cyclotron Road, Berkeley, CA 94720, USA}

\begin{abstract}

  A large population of ultra-diffuse galaxies (UDGs) was recently discovered in 
  the Coma cluster.  Here we present optical spectra of three such UDGs, DF~7, DF~44, 
  and DF~17, which have central surface brightnesses of $\mu_g \approx 24.4-25.1$ 
  mag arcsec$^{-2}$.  The spectra were acquired as part of an ancillary program 
  within the SDSS-IV MaNGA Survey.  We stacked 19 fibers in the central regions 
  from larger integral field units (IFUs) per source.  With over 13.5 hr of 
  on-source integration, we achieved a mean signal-to-noise ratio 
  (S/N) in the optical of $9.5$\AA$^{-1}$, $7.9$\AA$^{-1}$, and 
  $5.0$\AA$^{-1}$, respectively, for DF~7, DF~44, and DF~17.  Stellar 
  population models applied to these spectra enable measurements of 
  recession velocities, ages, and metallicities.  The recession velocities 
  of DF~7, DF~44, and DF~17 are $6599^{+40}_{-25}$~\kms, $6402^{+41}_{-39}$~\kms 
  and $8315^{+43}_{-43}$~\kms,  spectroscopically confirming that all 
  of them reside in the Coma cluster. The stellar populations of these three 
  galaxies are old and metal-poor, with ages of $7.9^{+3.6}_{-2.5}$~Gyr, 
  $8.9^{+4.3}_{-3.3}$~Gyr, and $9.1^{+3.9}_{-5.5}$~Gyr, and iron abundances of 
  $\mathrm{[Fe/H]}$ $-1.0^{+0.3}_{-0.4}$, $-1.3^{+0.4}_{-0.4}$ and 
  $-0.8^{+0.5}_{-0.5}$, respectively. 
  Their stellar masses are ($3$-$6$)$\times10^8$\solmass~.
  The UDGs in our sample are as old or older than galaxies 
  at similar stellar mass or velocity dispersion (only DF~44 has an independently 
  measured dispersion).  They all follow the well-established stellar 
  mass$--$stellar metallicity relation, while DF~44 lies below the velocity 
  dispersion-metallicity relation.  These results, combined with the fact that UDGs are unusually large 
  for their stellar masses, suggest that stellar mass plays a more important role in 
  setting stellar population properties for these galaxies than either size or surface brightness.
  
\end{abstract}
\section{Introduction}
 Spatially extended low surface brightness objects have 
 been found in galaxy cluster environments for decades 
 \citep{Sandage1984, Impey1988, Bothun1991, Dalcanton1997, Caldwell2006}. 
 Progress in identifying and characterizing such objects has been accelerated by specially 
 designed instruments and improved data reduction 
 methods in recent years.  For example, 
 \citet{vanDokkum2015} recently discovered a numerous population of low surface 
 brightness (${\mu}_{g} > 24$ mag arcsec$^2$) galaxies with surprisingly 
 large effective radii ($R_{\rm e} > 1.5$ kpc) in the Coma cluster 
 using the Dragonfly Telephoto Array \citep{Abraham2014}.  
 \citet{vanDokkum2015} referred to these objects as ultra-diffuse 
 galaxies (UDGs). 
 \footnote{Note the term UDG is not universally used to describe all spatially extended 
and low surface brightness objects.}

\vskip 0.1cm 
\begin{figure*}[t] 
\centering 
\includegraphics[width=18.0cm]{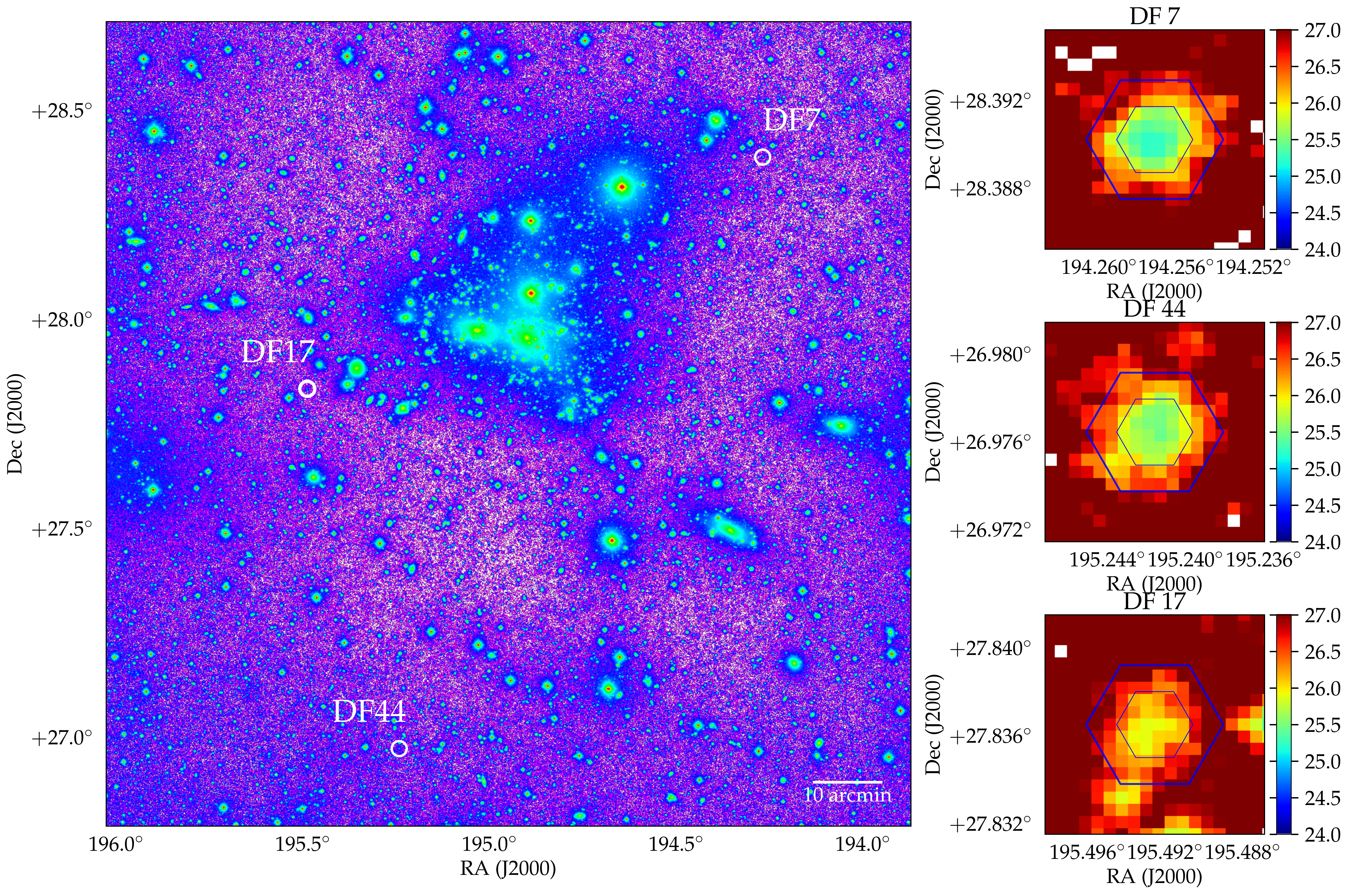}
\caption{
  Left panel: overview of the MaNGA fiber bundle locations of three UDGs on 
  the g-band surface brightness map of the Coma cluster.  The image is from the 
  Dragonfly Telephoto Array.  Three white circles show the locations of DF~7, 
  DF~44, and DF~17 but are not true to the IFU size; 
  Right panels: zoomed-in $g$-band surface brightness map of three UDGs.  
  Large hexagons show the footprints of MaNGA 61-fiber IFUs, and small hexagons 
  show the stacked regions (19 fibers).  Colorbars show the surface brightness 
  in the $g$-band.}
\label{locfig}
\end{figure*}
  
 Many UDGs are found to have exponential-like surface brightness 
 profiles, and axis ratios similar to dwarf spheroidal galaxies 
 \citep[e.g.][]{vanDokkum2015, Koda2015}.
 Most UDGs in cluster environments appear to have colors that are 
 consistent with the low-mass end of the red-sequence, indicating 
 that they are indeed cluster members.  As their name implies, 
 their sizes are much larger compared to dwarf elliptical galaxies.  
 Following \citet{vanDokkum2015},  more UDGs have been found in the 
 Coma cluster.  For instance, \citet{Koda2015} found more than 
 800~UDGs in the Coma cluster based on imaging from the Suprime--Cam 
 on the Subaru telescope. More than 300 of these UDGs have 
 $R_{\rm e} > 1.5$ kpc \citep[also see][]{Yagi2016}.  
 UDGs have also been found in other nearby galaxy clusters 
 \citep[e.g.,][]{Mihos2015, Munoz2015, MartinezDelgado2016, Roman2017, 
 vanderBurg2016}, group environments \citep[e.g.,][]{Merritt2016} 
 and the field \citep[e.g.,][]{Bellazzini2017, Leisman2017}. 
 Owing to their diffuse nature, only a few UDGs have a measured spectroscopic redshift 
 \citep[e.g.,][]{vanDokkum2015, Kadowaki2017}. 

 Furthermore, some UDGs show evidence for being associated with large 
 numbers of globular clusters \citep[GCs;][]{Beasley2016a, Beasley2016b, 
 Peng2016, vanDokkum2016, vanDokkum2017}.  For example, DF~17 
 ($\mathrm{M_g=-15.2}$) in the Coma cluster has $\approx 30$ GCs 
 \citep{Peng2016}, and DF~44 ($\mathrm{M_g=-15.7}$) is 
 associated with $\approx 70$ GCs 
 \citep{vanDokkum2016,vanDokkum2017}. In addition, DF~44 was revealed 
 to have a surprisingly high velocity dispersion for its stellar mass
 \citep[$\sigma=47^{+8}_{-6}$~km/s, ][]{vanDokkum2016}.  
 Both the high velocity dispersion and the rich GC 
 population suggest that at least some UDGs live in relatively massive dark 
 matter halos 
 \citep[e.g., $\mathrm{M_{tot, DF~44}\sim 10^{12}M_\odot}$, ][]{vanDokkum2016}.

 There is not yet a clear picture for the formation of UDGs. One plausible 
 formation scenario is that UDGs such as DF~44 are failed Milky Way-like 
 galaxies that have truncated star formation very early on 
 \citep[e.g.,][]{vanDokkum2015b,vanDokkum2016,vanDokkum2017}, due to strong 
 feedback or environment-related processes.  Under this scenario, UDGs 
 should be outliers of the stellar mass--halo mass relation and have more 
 massive dark matter halos for their stellar masses.  This is supported by 
 the high stellar velocity dispersion and the rich GC systems 
 associated with them.  Other scenarios include structural transformations via tidal 
 interaction in the cluster environment \citep[e.g.,][]{Collins2013,Merritt2016}, 
 or formation within dark matter halos with unusually high angular momenta \citep{Amorisco2016}.   

 While the formation of UDGs is still a puzzle, stellar population 
 information should provide useful insights. For example, low mass galaxies 
 reside on a tight stellar mass--stellar metallicity relation 
 \citep[e.g.,][]{Kirby2013}, and this relation seems to be continuous to 
  high masses \citep[e.g.,][]{Gallazzi2005, Kirby2013, Conroy2014}.  
 Previous work suggests that this apparently tight relation is linked to 
 multiple effects, such as metal loss due to supernova ejecta, star 
 formation efficiency and gas inflows \citep[e.g.,][]{Ma2015, Lu2017}.  
 Interestingly, the special properties of UDGs such as their diffuse 
 structures and relatively massive dark matter halos may also be related to the above processes.
 The low stellar surface density of UDGs may imply low star formation 
 efficiency. The massive dark matter halos may imply that it is harder 
 for UDGs to lose gas and metals due to the deep gravitational 
 potential, but at the same time the metals in the interstellar 
 medium may be more easily diluted. Whether UDGs follow the same stellar 
 mass--metallicity relation as low mass galaxies can provide us 
 with more clues on how the above processes jointly act. 

 In this paper, we present the first stellar population analysis through 
 full spectral modeling for three UDGs, DF~7, DF~44 and DF~17, in the Coma cluster 
 based on data obtained as part of an ancillary program within the 
 SDSS-IV/MaNGA program.  The Coma cluster has a median redshift of 
 $cz=7090$~\kms \citep{Geller1999} and a velocity dispersion 
 $\sim 1000$~\kms \citep[e.g.,][]{Colless1996, Mobasher2001, Rines2013, 
 Sohn2016}.  All of the three UDGs have large sizes for their stellar mass
 \citep[$\mathrm{R_{eff, DF~7}=4.3^{+1.4}_{-0.8}}$~kpc, 
 $\mathrm{R_{eff, DF~44}=4.6^{+1.5}_{-0.8}}$~kpc, 
 $\mathrm{R_{eff, DF~17}=4.4^{+1.5}_{-0.9}}$~kpc,][]{vanDokkum2015}, indicating they are 
 typical examples in the UDG population. We provide the first 
 measurement of a recession velocity of DF~17, and confirmation 
 measurements of the recession velocities of DF~7 and DF~44.  
 We also present the first measurement of their age and metallicity.  
 The distance of the Coma Cluster is assumed to be 100 Mpc, which was 
 adopted from \citet{Liu2001}. This corresponds to a distance modulus of 34.99 
 mag and a scale of 0.474 kpc arcsec$^{-1}$.  The Galactic extinction 
 for the Coma Cluster is $A_{\rm SDSS-g}=0.030$ mag and 
 $A_{\rm SDSS-r}=0.021$ mag \citep{Schlafly2011}, which is 
 applied to the photometry in this paper.  
 All magnitudes given in this paper are in the AB system.

\section{Data}

\subsection{Project Overview} 
 We make use of data obtained by the MaNGA Survey (Mapping Nearby Galaxies 
 at Apache Point Observatory, \citep{Bundy2015, Yan2016b, Drory2015}).  
 MaNGA is a large, optical integral field spectroscopy survey with 17 
 deployable integral field units (IFUs) (ranging from 12\asec 
 to 32\asec in diameter ), and 
 one of the fourth-generation Sloan Digital Sky Survey (SDSS-IV) programs 
 \citep{Blanton2017}.  The primary goal of MaNGA is to obtain an integral field 
 spectroscopy of $\sim10,000$ nearby galaxies.  

Our data comes from one of MaNGA's ancillary programs, the Deep Coma 
program
\footnote{http://www.sdss.org/dr14/manga/manga-target-selection/ancillary-targets/coma/}.  
The goal of the Deep Coma program is to study the stellar populations 
of various targets in the Coma Cluster and its surrounding area through long integration spectroscopy.  
Our targets include two brightest cluster galaxies (BCG), several additional 
massive elliptical galaxies, dwarf galaxies, intracluster light regions 
and three UDGs, DF~44, DF~17 and DF~7. 

\subsection{Observations}
 MaNGA makes use of IFU fiber bundles to feed two dual-beam Baryonic 
 Oscillation Spectroscopic Survey (BOSS) spectrographs 
 \citep{Smee2013, Drory2015} that are on the SDSS 2.5 meter telescope 
 \citep{Gunn2006}. The spectrographs have 1423 fibers in total that are 
 bundled into different size IFUs. The diameter of each fiber is 
 $1.98$\asec~ on the sky. The wavelength coverage of the spectrographs 
 is $3622 –- 10354$\AA~ with a $\sim 400$\AA~ overlap from 
 $\sim5900$\AA~ to $\sim6300$\AA~.  The spectral resolution is $1560 –- 2650$. 
 
The Deep Coma project consists of six plates designed to observe 
selected targets in the Coma cluster.  The centers of all plates are at 
RA$=12^h58^m35^s.58$, DEC$=27^d36^m12^s.744$.  This position was carefully 
chosen to optimize the IFU bundle mapping of desired targets. 
The first two plates were observed in Spring, 2015.  
The third to fifth plates were observed in Spring, 2016.
The sixth plate was observed in Spring, 2017.
The total on-source exposure time in each plate is 2.25 hr. 
To evaluate the impact on low surface brightness targets from systematic 
residuals (see Section~$2.3$), we have shuffled the IFU bundles used for 
the same target among different plates.  The most frequently used IFU 
bundles for UDGs are 61-fiber bundles, but 37-fiber bundles and 
91-fiber bundles have been used to observe them as well.
The stacked spectra in our analyses are from the inner 19 fibers of 
each bundle for the optimum integrated signal-to-noise ratio (S/N). 
The locations of the stacked region for the three UDGs 
analyzed in this paper are shown in Figure~\ref{locfig}.  
The diameter of each stacked region on the sky is $12$\asec$.5$.  
The positions, central surface brightness, and effective radii of the two 
UDGs are shown in Table~1. The stacked regions sample roughly the 
inner $0.7$~effective radii of the three UDGs.

To probe low surface brightness regions, excellent sky subtraction is 
required. In the Deep Coma plates, the locations of the reference sky 
fibers are carefully selected using images from the Dragonfly 
Telephoto Array. The expected $g$-band surface brightness of all sky 
fiber locations ($\sim 3$~arcsec around each fiber) based on the 
Dragonfly imaging are $\mathrm {> 27.8}$ $\mathrm {mag/{arcsec^2}}$.  
In addition to the 92 single fibers used to construct the model sky 
spectrum for ordinary MaNGA plates, we also devote three IFU bundles 
(two 19-fiber bundles, one 37-fiber bundle) to additional measurements 
of the sky. These additional sky fibers enable us to reach fainter 
surface brightness limits for stacked spectra than the regular MaNGA 
survey.  The $1 \sigma$ limiting surface brightness we are able to 
detect is calculated from the $1 \sigma$ rms in the wavelength range 
$4000$--$5500$~\AA,  and is 27.6 \sb~ for the Deep Coma plates.

\begin{deluxetable}{lccc}
\tablecaption{UDG Properties}
\tabletypesize{\scriptsize}
\tablehead{
\colhead{Target} &
\colhead{DF~7} &
\colhead{DF~44} &
\colhead{DF~17}
}
\startdata
\begin{tabular}{l}$\alpha$\end{tabular}& $12^h57^m01^s.7$ & $13^h00^m58^s.0$ & $13^h01^m58^s.3$ \\
\noalign{\smallskip}\noalign{\smallskip}
\begin{tabular}{l}$\delta$\end{tabular}& $28^{\circ}23^{\prime}25^{\prime\prime}$ & $26^{\circ}58^{\prime}35^{\prime\prime}$ & $27^{\circ}50^{\prime}11^{\prime\prime}$\\
\noalign{\smallskip}\noalign{\smallskip}
\begin{tabular}{l}$\mu_{0,g}$(mag arcsec$^{-2}$)\end{tabular}&
$24.4\pm0.5$ & $24.5\pm0.5$ & $25.1\pm0.5$   \\
\noalign{\smallskip}\noalign{\smallskip}
\begin{tabular}{l}$M_{g}$(mag)\end{tabular}&
$-16.0^{+0.2}_{-0.2}$ & $-15.7^{+0.2}_{-0.2}$ & $-15.2^{+0.3}_{-0.2}$   \\
\noalign{\smallskip}\noalign{\smallskip}
\begin{tabular}{l}$R_{\rm eff}$(kpc)\end{tabular}&
$4.3^{+1.4}_{-0.8}$ & $4.6^{+1.5}_{-0.8}$ & $4.4^{+1.5}_{-0.9}$ \\
\noalign{\smallskip}\noalign{\smallskip}
\begin{tabular}{l}S$/$N($4500-5000$\AA)\end{tabular}& 
$9.5$\AA$^{-1}$ & $7.9$\AA$^{-1}$ & $5.0$\AA$^{-1}$\\
\noalign{\smallskip}\hline\noalign{\smallskip}
\multicolumn{4}{c}{Constraints from Spectra} \\
\noalign{\smallskip}\hline\noalign{\smallskip}
\begin{tabular}{l}Velocity (\kms)\end{tabular}&
$6600^{+40}_{-26}$ & $6402^{+41}_{-38}$ & $8311^{+43}_{-43}$  \\
\noalign{\smallskip}\noalign{\smallskip}
\begin{tabular}{l}log(age/Gyr)\end{tabular}& $0.93^{+0.17}_{-0.18}$ & $1.02^{+0.11}_{-0.24}$ & $0.88^{+0.22}_{-0.42}$  \\
\noalign{\smallskip}\noalign{\smallskip}
\begin{tabular}{l}[Fe/H]\end{tabular}& $-1.03^{+0.31}_{-0.34}$ & $-1.25^{+0.33}_{-0.39}$ & $-0.83^{+0.56}_{-0.51}$\\
\noalign{\smallskip}\noalign{\smallskip}
\begin{tabular}{l}$(M/L)_r$\end{tabular}& $1.63^{+0.55}_{-0.29}$ & $1.86^{+0.39}_{-0.56}$ & $1.54^{+0.71}_{-0.52}$\\
\noalign{\smallskip}\noalign{\smallskip}
\begin{tabular}{l}$\mathrm{\log(M_\star/M_\odot)}$\end{tabular}& $8.74^{+0.17}_{-0.11}$ & $8.66^{+0.12}_{-0.15}$ & $8.42^{+0.22}_{-0.19}$\\
\noalign{\smallskip}\hline\noalign{\smallskip}
\multicolumn{4}{c}{Combined Constraints from Spectra and Photometry} \\
\noalign{\smallskip}\hline\noalign{\smallskip}
\begin{tabular}{c}Velocity (\kms)\end{tabular}&
$6599^{+40}_{-25}$ & $6402^{+41}_{-39}$ & $8315^{+43}_{-43}$  \\
\noalign{\smallskip}\noalign{\smallskip}
\begin{tabular}{l}log(age/Gyr)\end{tabular}& $0.90^{+0.17}_{-0.16}$ & $0.95^{+0.17}_{-0.20}$ & $0.96^{+0.16}_{-0.40}$ \\
\noalign{\smallskip}\noalign{\smallskip}
\begin{tabular}{l}[Fe/H]\end{tabular}& $-1.04^{+0.32}_{-0.36}$  &  $-1.25^{+0.35}_{-0.41}$ &  $-0.80^{+0.49}_{-0.47}$\\
\noalign{\smallskip}\noalign{\smallskip}
\begin{tabular}{l}$(M/L)_r$\end{tabular}& $1.56_{-0.28}^{+0.47}$ & $1.64^{+0.54}_{-0.38}$ & $1.80^{+0.51}_{-0.66}$\\
\noalign{\smallskip}\noalign{\smallskip}
\begin{tabular}{l}$\mathrm{\log(M_\star/M_\odot)}$\end{tabular}& $8.72_{-0.13}^{+0.17}$ & $8.61_{-0.11}^{+0.16}$ & $8.49_{-0.20}^{+0.15}$\\
\enddata
\tablecomments{We adopt the locations, central $g$-band surface brightness, absolute $g$-band 
magnitude and effective radii from \citet{vanDokkum2015}. Stellar population properties are 
derived from the central 19 fibers.}
\end{deluxetable}

To further improve the accuracy of the background estimate and to mitigate 
systematics, we adopt an on-and-off nodding strategy. The goal is to obtain 
reference `all-sky' exposures by shifting the whole field approximately $20^{\prime}$ away, 
so that most of the sky fibers and science IFUs will sample blank sky.
We searched for the ideal locations for the shifts using images from the Dragonfly 
Telephoto Array to optimize the fraction of both the science and sky fibers 
on the low surface brightness region. 
Each of the first two plates includes nine 5-minute 
nodding exposures at nine different locations between the normal science 
exposures. Each of the last four plates includes four 15-minute nodding 
exposures at four different locations.  

\begin{figure}[t]
\centering 
\includegraphics[width=8.5cm]{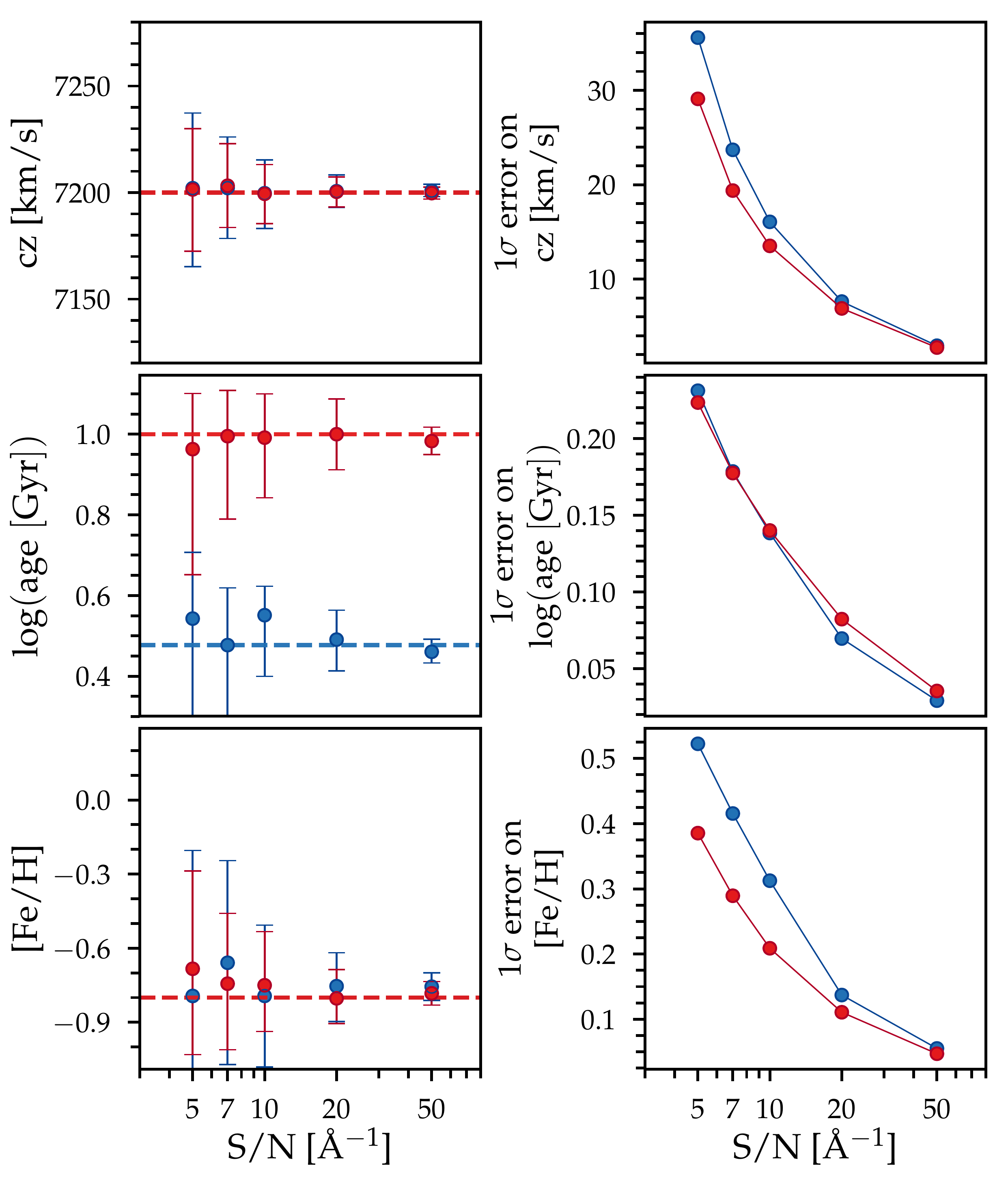}
\caption{
 Test of the recovery of recession velocity, age and [Fe/H] with 
 mock spectra as a function of S/N. We constructed 20 realizations at S/N 
 of 5, 7, 10, 20 and $50$\AA$^{-1}$.  
 The different colors represent mock spectra with different stellar ages, 
 where the blue and red represent model with age of $3$~Gyr and $10$~Gyr, respectively.
 Left panels: true values are shown as horizontal dashed lines. Circles, 
 lower and upper error bars show the mean values of 50 th, 16 th and 84 th percentiles 
 of the posterior distributions of 20 realizations.
 Right panels:  $1\sigma$ uncertainties of recession velocity, 
 log(age/Gyr) and [Fe/H] as a function of S/N, averaged over the 20 realizations.
 }
\label{mock}
\end{figure}
\noindent  

\subsection{Data Reduction}

We processed our data using version 2\_2\_0 of the MaNGA Data Reduction Pipeline
\citep[DRP;][]{Law2015, Law2016}.  This version is similar to v2\_1\_2 released 
in 2017 July as SDSS DR14 \citep{Abolfathi2017}, but incorporates a number
of custom modifications made specifically for the Deep Coma program.
The baseline DRP first removes detector overscan regions and quadrant-dependent bias 
and extracts the spectrum of each fiber using an optimal profile-fitting technique.  
It next uses the sky fibers to create a super-sampled model for the background sky 
spectrum and subtracts this model spectrum from each of the science fibers.  Flux 
calibration is then performed on individual exposures using 12 7-fiber IFUs targeting 
spectrophotometric standard stars \citep{Yan2016a}.  Fiber spectra from the blue and 
red cameras are then combined together onto a common logarithmic wavelength solution 
using a cubic b-spline fit.  These `mgCFrame' files thus represent spectra of all 
1423 MaNGA fibers from a single exposure in a row-stacked format, where
each row corresponds to an individual 1-dimensional fiber spectrum.
The logarithmic wavelength grid runs from 
$\log{\lambda}$(\AA)$=3.5589$ to $\log{\lambda}$(\AA)$=4.0151$, 
which corresponds to 4563 spectral elements from 
3621.5960 to 10353.805 \AA.
We do not utilize the 3d stage DRP (which combines fiber spectra
into 3-dimensional data cubes), and restrict our analysis to
the mgCFrame row-stacked spectra.

A number of custom modifications were made to the DRP in v2\_2\_0
to optimize performance for observations of low 
surface brightness regions in the Coma cluster.  First, the DRP
was modified to use the 167 total sky fibers (92 single fibers
plus 3 IFU fiber bundles) to construct the model sky spectrum.
Secondly, analysis of our nodded all-sky observations showed evidence
for low-level systematics in the detector electronics.  We added
a step in the pipeline to measure and remove a 0.5 e-/pixel offset 
in bias between the light-sensitive detector pixels and the 
overscan region, compensating at the same time for a 
seasonally-dependent 0.1 electron/pixel drift in the difference.
Additionally, we found that the amplifier-dependent gain values
tended to drift from one exposure to the next away from nominal
at the $\sim0.1\%$ level; we added procedures to measure and 
correct for this effect empirically using the sky fibers in each 
exposure.  Finally, we modified the DRP to be able to apply the 
flux calibration vector from the nearest (in time) ordinary science 
exposure to the nod exposures (for which there are no calibration 
stars in the 7-fiber mini bundles). 

Additionally, performance analysis of early observations in the Deep
Coma program revealed that scattered light and the extended 
($>100$ pixel) profile wings of bright galaxies targeted by the Coma 
program were contaminating the spectra of fainter objects.  We therefore 
redesigned our observing program to consolidate all bright targets 
(central and dE galaxies) onto one of the two BOSS spectrographs, and 
all faint targets (UDGs and intracluster light) onto the other so that 
these targets never share a detector.  

Although these modifications substantially improve performance for the 
Deep Coma program relative to the DR14 baseline DRP, we find that the
final stacked science spectra are nonetheless still limited
by systematic residuals over large wavelength scales ($>100$\AA).
These residuals are consistent between stacked science and nodded 
sky spectra within each plate, possibly due to cartridge-dependent 
uncertainties in fiber alignment and the detector point-spread function.  
For example, in the fifth plate, the stacked science spectrum for DF~17 
is systematically negative in flux, but the nodded sky spectrum 
is systematically more negative, and the offset ranges from 0 to 
10$^{-18}$erg~s$^{-1}$cm$^{-2}$\AA$^{-1}$.
In the last four plates, we mitigate 
the impact of these systematics by fitting the stacked spectra of the 
sky subtracted nodded sky exposures with a tenth degree polynomial from 
3836 to 5873 \AA in the observed frame and subtracting 
the result from the corresponding science exposures prior to stacking 
science spectra.   The amplitude of polynomial correction ranges 
from $10^{-19}$ to $10^{-18}$erg~s$^{-1}$cm$^{-2}$\AA$^{-1}$ in the 
continuum, and represents an important correction to the baseline 
flux level for extremely faint targets.
We manage to match the continuum levels of stacked spectra from 
different plates via subtracting the above polynomial continuum 
before we derive any science result.

The mean signal-to-noise ratio (S/N) we achieved after all of these steps was
$9.5$\AA$^{-1}$ for DF~7, $7.9$\AA$^{-1}$ for DF~44 and $5.0$\AA$^{-1}$ for DF~17 
in the observed wavelength range $4500-5000$\AA.  This corresponds to a 
total integration time on source of 13.5 hr, and a total integration 
time on nod exposures of 4 hr.

\begin{figure*}[t] 
\centering 
\includegraphics[width=16.5cm]{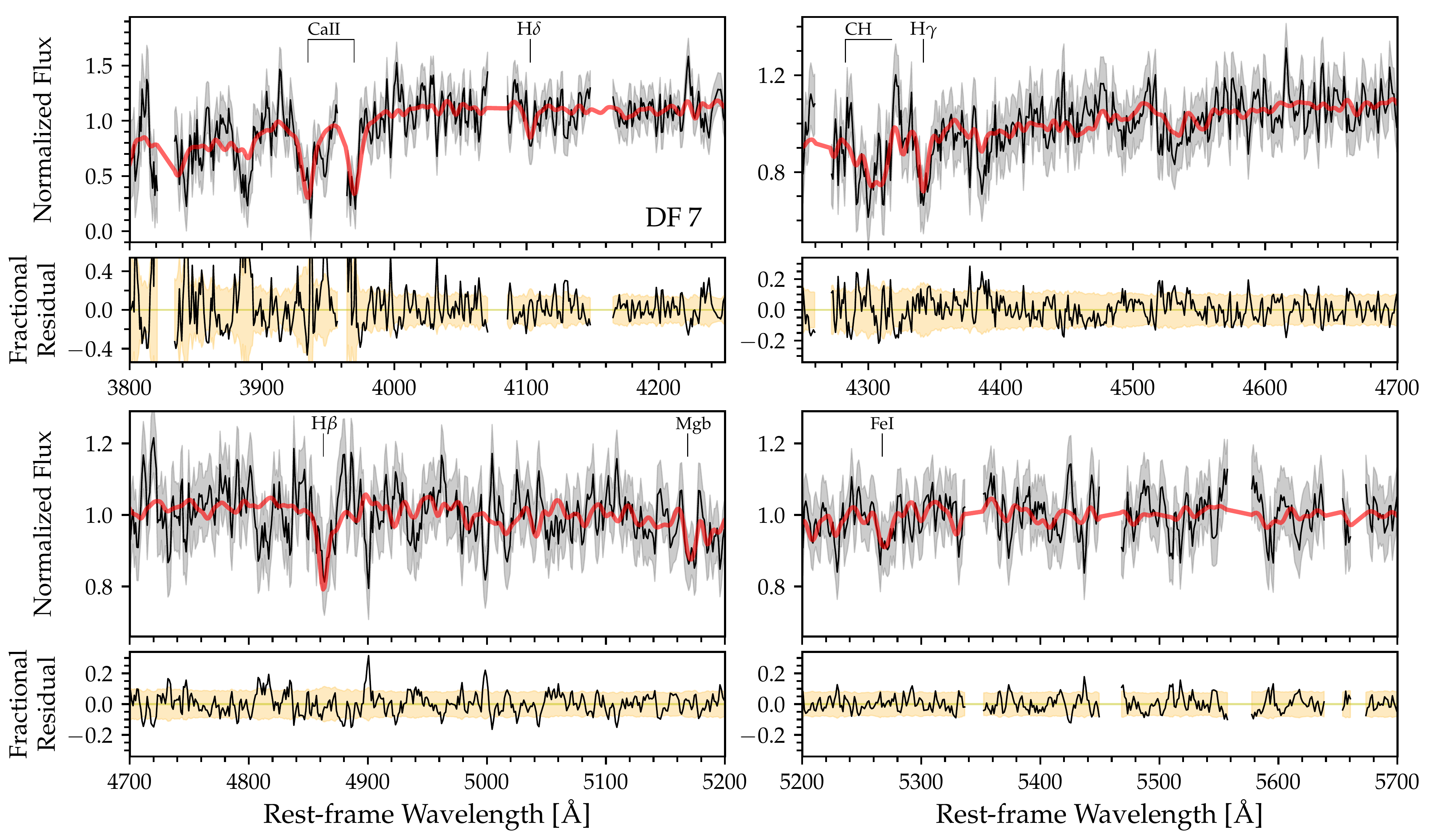}
\caption{
 Top panel: Stacked spectrum of DF~7 (black) and best-fit model 
 spectrum (red) with parameters at minimum $\chi^2$ from {\tt alf}.
 Gray shaded regions show the uncertainty of flux from 
 the input spectrum.
 Bottom panel: Fractional residuals (black) are compared with 
 the uncertainty of flux from the input spectrum (yellow).
 Gaps on the black lines indicate pixels that are masked prior to the 
 fitting, which are pixels under the bright sky lines.
}
\label{spec1}
\end{figure*}
\begin{figure}[b]
\centering 
\includegraphics[width=7.5cm]{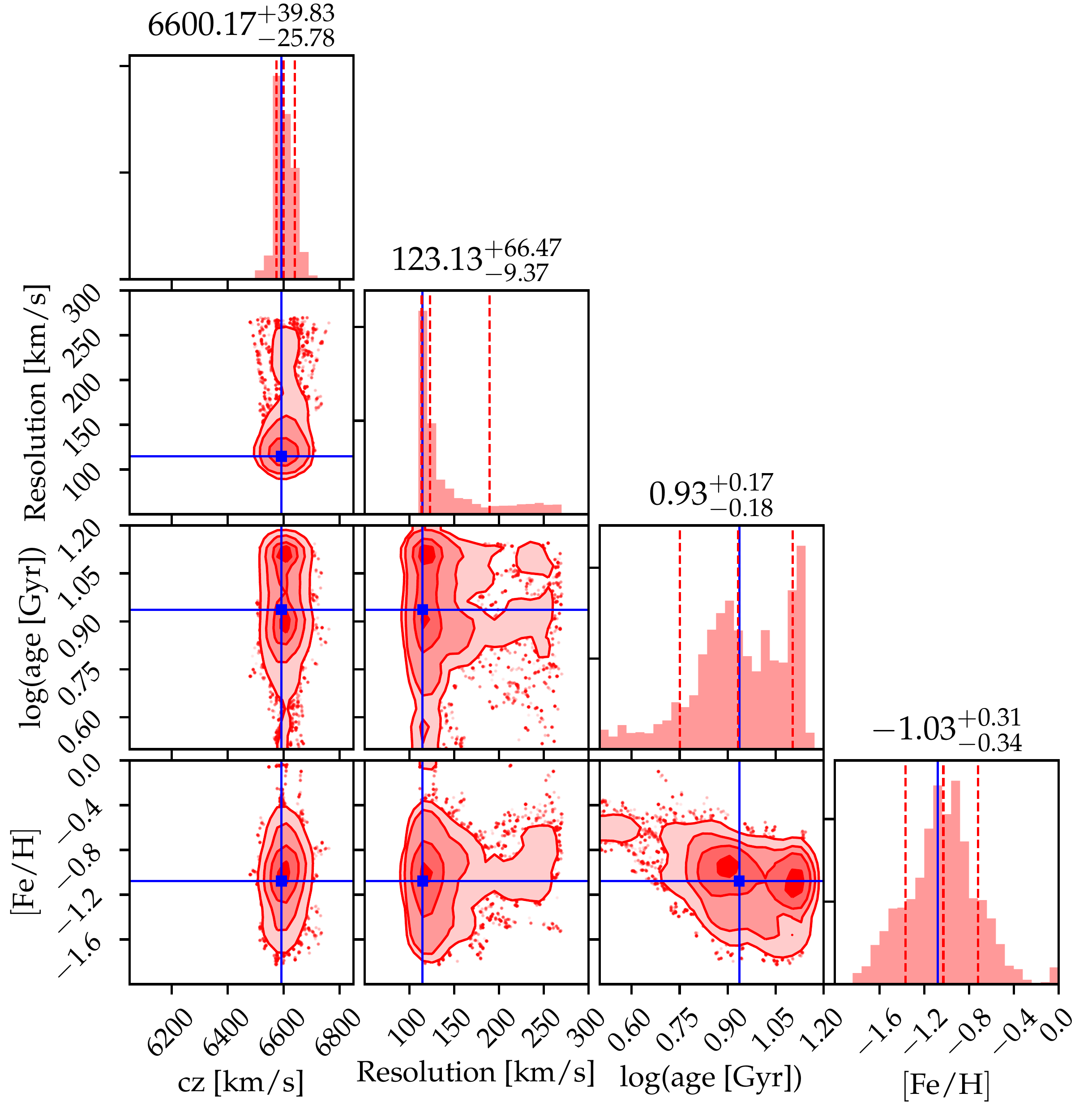}
\caption{
 Projections of the posterior of recession velocity, log(age) and [Fe/H] 
 from {\tt alf} in 1D and 2D histograms for DF~7. Dashed lines in 1D 
 histograms and contours in 2D histograms show the 16th, 50th and 84th 
 percentiles of posteriors. Blue lines represent the best fit parameters 
 at minimum $\chi^2$, which are used to generated best-fit model spectra.
}
\label{posterior1}
\end{figure}

\section{Stellar Population Modeling}

\subsection{Absorption Line Fitter}
 Our main tool for modeling spectra of galaxies and UDGs in our sample 
 is the absorption line fitter \citep[{\tt alf},][]{Conroy2012, Conroy2014, 
 Conroy2016}.   {\tt alf} enables stellar population modeling of the full 
 spectrum for stellar ages $>1$Gyr and for metallicities from $\sim-2.0$ to $+0.25$.  
 With {\tt alf} we explore the parameter space using a Markov Chain Monte Carlo 
 algorithm \citep[{\tt emcee},][]{ForemanMackey2013}.  The program now adopts 
 the MIST stellar isochrones \citep{Choi2016} and utilizes a new spectral library 
 that includes continuous wavelength coverage from $0.35-2.4\mu m$ over a wide 
 range in metallicity.  This new library, described in \citet{Villaume2017}, 
 is the result of obtaining new IRTF NIR spectra for stars in the MILES 
 optical spectral library \citep{SanchezBlazquez2006}.  Finally, theoretical 
 response functions, which tabulate the effect on the spectrum of 
 enhancing each of 18 individual elements, were computed using the 
 ATLAS and SYNTHE programs \citep{Kurucz1970, Kurucz1993}.  Further details of 
 these updates to {\tt alf} are described in \citet{Conroy2018}.
 With {\tt alf} we are able to fit a two burst star formation history, 
 the redshift, velocity dispersion, overall metallicity ([Z/H]), 18 
 individual element abundances, several IMF parameters, and a variety 
 of ``nuisance'' parameters.
 
\begin{figure*}[t] 
\centering 
\includegraphics[width=16.5cm]{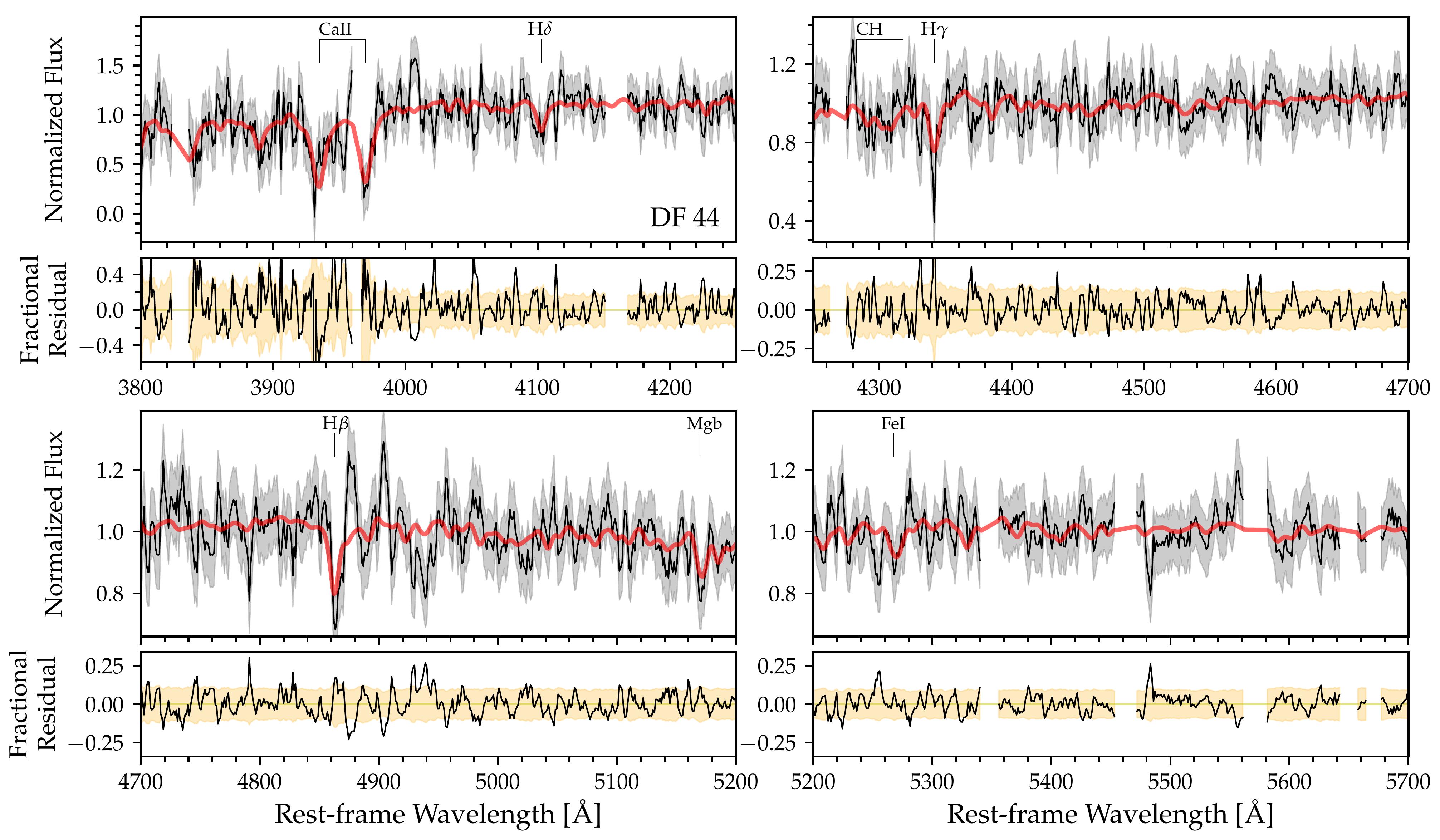}
\caption{
Same as Figure~\ref{spec1}, but for DF~44. 
}
\label{spec2}
\end{figure*}
\vskip 0.25cm
\begin{figure}[b]
\centering 
\includegraphics[width=7.5cm]{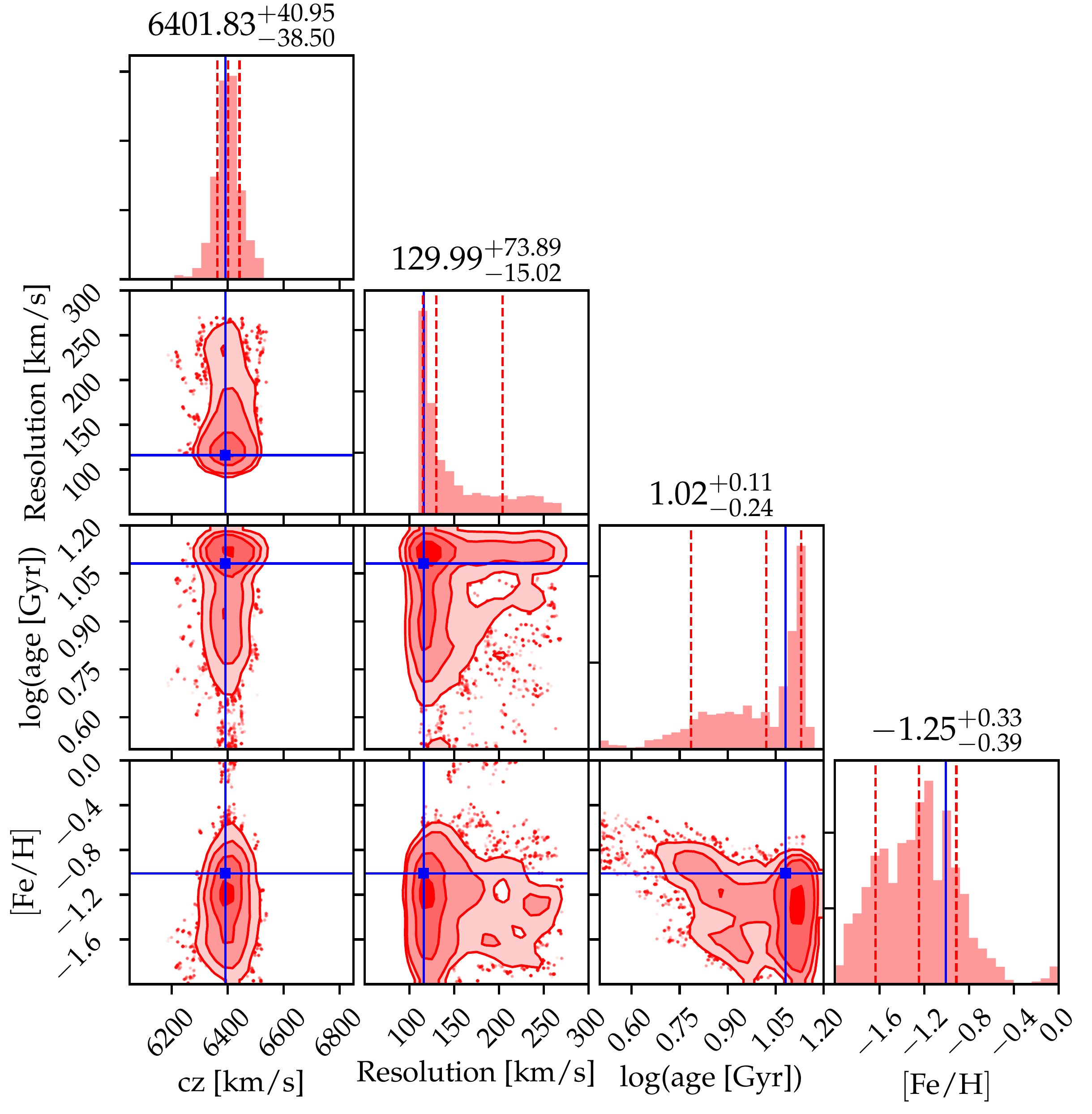}
\caption{
Same as \ref{posterior1}, but for DF~44. 
}
\label{posterior2}
\end{figure}
\vskip 0.25cm
\noindent  

 Throughout this paper, we use {\tt alf} in a simplified mode.  
 Not all the parameters are included, but only the recession velocity, 
 age, overall metallicity [Z/H] and abundances of ~Fe, ~C, ~N, ~O, ~Mg, 
 ~Si, ~Ca, ~Ti, and Na.  
 The IMF is fixed to the \citet{Kroupa2001} form.  
 Instead of adopting a two burst star formation history in the standard 
 model, the simplified mode adopts only a single age component.
 We adopt this approach due to the limited S/N of the data.  
 The intrinsic velocity dispersions of UDGs \citep{vanDokkum2016} are 
 assumed to be 
 lower than both the instrumental resolution \citep{Law2016} and the 
 resolution of models. Therefore we smoothed each spectrum  based on 
 the instrumental resolution of the corresponding fiber. The desired 
 velocity resolutions ($\sigma_D$) are chosen based on the maximum 
 instrumental resolution (maximum $\sigma_i$), and are different among 
 three galaxies.  The desired velocity resolutions are 110~km/s for DF~7, 
 125~km/s for DF~44 and 135~km/s for DF~17.  
 Each spectrum is smoothed to the desired velocity resolution by 
 convolving a wavelength dependent Gaussian kernel with 
 $\sigma=\sqrt{{\sigma_D}^2-{\sigma_i}^2}$.  With {\tt alf} we fit for 
 the velocity dispersion of the smoothed spectra and describe this 
 property below as `resolution'.  We adopt flat priors from 
 $500-10500$~km/s for recession velocity, $10-500$~km/s for resolution, 
 $1.0-14$~Gyr for age and $-1.8-+0.3$ for [Fe/H].  The priors are zero 
 outside these ranges. For each spectrum we fit a continuum in the 
 form of a polynomial to the ratio between model and data.  The polynomial 
 order is $(\lambda_{max}$--$\lambda_{min})/100$\AA.  
 During each likelihood call the polynomial divided input spectrum and 
 model are matched.  For computational convenience the continuum 
 normalization occurs in two separate wavelength intervals, 
 $3800-4700$\AA~ and $4700-5700$\AA~\citep{Conroy2012b}.  
 In this paper we only use the data 
 taken by the blue spectrograph.  This allows us to avoid additional issues 
 associated with the numerous bright atmospheric OH features in the red.  
 Pixels near bright sky lines in the blue were masked prior to the fitting.

\begin{figure*}[t] 
\centering 
\includegraphics[width=16.5cm]{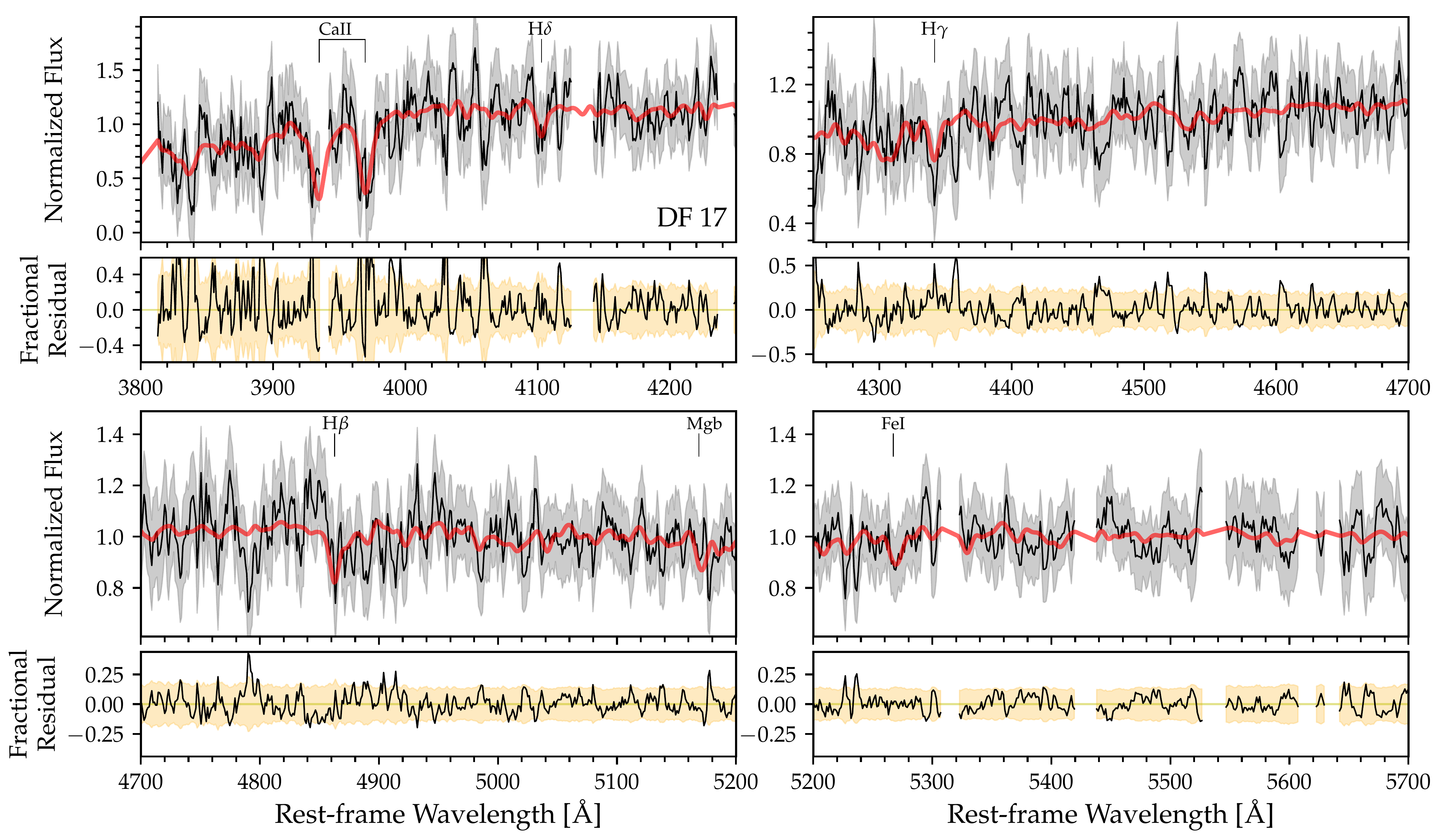}
\caption{
Same as Figure~\ref{spec1}, but for DF~17. 
}
\label{spec3}
\end{figure*}
\vskip 0.25cm
\begin{figure}[b]
\centering 
\includegraphics[width=7.5cm]{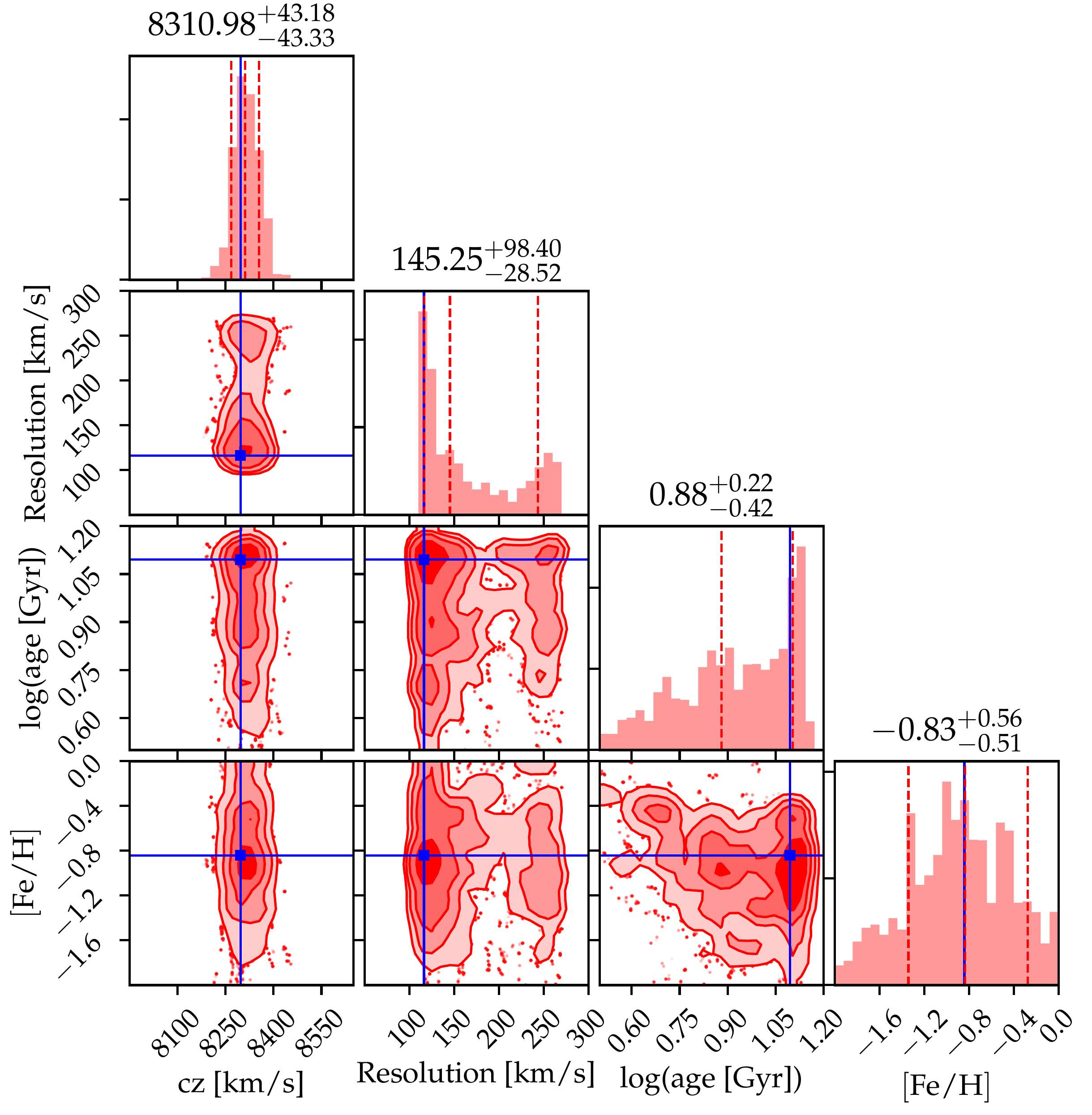}
\caption{
Same as \ref{posterior1}, but for DF~17. 
}
\label{posterior3}
\end{figure}
\vskip 0.25cm
\noindent  
\subsection{Mock Data Tests}

 In this section, we examine how well the recession velocity, age and 
 [Fe/H] can be recovered with {\tt alf}.  We have constructed a mock 
 spectra dataset with 20 realizations at a range of S/N from 5 to 
 $50$\AA$^{-1}$.  We assume that the mock spectra have a true recession velocity 
 of $7200$~\kms, a velocity dispersion of $50$~\kms 
 and a metallicity of [Fe/H]~$=-0.8$. We test a set of mock spectra 
 with ages of $10$ Gyr and the other with ages of $3$ Gyr.
 The data are fit over two wavelength ranges,  $3800-4700$\AA~ and $4700-5700$\AA.  
 The fitting ranges are the same as what we apply to the UDG spectra.  
 In addition, the mock spectra are convolved by a $100$~\kms Gaussian kernel 
 since we need to smooth the UDG spectra. The results are shown in 
 Figure~\ref{mock} as a function of S/N.  

 The S/N of our UDG spectra lie between 5 and 10.  From the mock test 
 of the old stellar population, we 
 estimate that their recession velocities, log(age/Gyr), and [Fe/H] can be 
 reliably measured with an uncertainty of $14-29$~\kms, $0.14-0.22$ and 
 $0.21-0.39$~dex, respectively.  
 From the mock test of the $3$ Gyr stellar population, we 
 estimate that their recession velocities, log(age/Gyr) and [Fe/H] can be 
 reliably measured with an uncertainty of $16-36$~\kms, $0.14-0.23$ and 
 $0.31-0.52$~dex, respectively.  The recession velocity can be recovered 
 at good accuracy for mock spectra at all S/N.   The bias in recovered 
 parameter is small even at low S/N: for log(age/Gyr) and 
 [Fe/H] the bias is of the order $0.01-0.04$ and $0.05-0.12$~dex, well 
 within the statistical uncertainties.  

\section{Results}

 We now present our results from full spectral modeling.  A summary of the 
 available information of the three targets are shown in Table~1, including 
 their locations, $g$-band central surface brightness, effective radii 
 \citep{vanDokkum2014a}, mean S/N of the spectra, and derived stellar population 
 properties.

\subsection{Stellar Populations for DF~7, DF~44, and DF~17}

 In this section, we present our constraints on the redshift, age,  
 and metallicity for DF~7 (Figure~\ref{spec1} and \ref{posterior1}), 
 DF~44 (Figure~\ref{spec2} and \ref{posterior2}), and 
 DF~17 (Figure~\ref{spec3} and \ref{posterior3}). For each parameter, 
 we present the median values and the 16th and 84th percentile of 
 the posterior distributions.  Top panels of Figure~\ref{spec1}, 
 \ref{spec2}, and \ref{spec3} show the normalized median 
 stacked spectra of DF~7, DF~44, and DF~17 with the inner 19 fibers over 
 all exposures in black, and our best-fit model spectra in red. 
 The best-fit model spectra are generated with parameters at the minimum $\chi^2$.
 
 The bottom panels of Figure~\ref{spec1}, \ref{spec2}, and \ref{spec3} 
 show the fractional residuals in black, compared 
 with flux uncertainty in the yellow shaded regions.  The comparison 
 between the residual and data uncertainty indicates that the fitting 
 results for all three UDGs are successful, as the residuals 
 are all consistent with the flux uncertainty.  
 For DF~7 and DF~44, several prominent absorption line features that are 
 well-known for estimating stellar age and metallicity, such as H$\delta$, 
 H$\gamma$, H$\beta$, Mg {\it b} and the G-band at $\sim4300$\AA~ are all 
 well described by the best-fit model.  
 The Ca II H \& K lines are very prominent as well.  The spectrum 
 for DF~17 has a much lower S/N, but key features such as H$\gamma$, H$\beta$ 
 are all clearly visible and well-described by the best fit model. 

 In Figure~\ref{posterior1}, \ref{posterior2}, and \ref{posterior3}, 
 we show the projections of posteriors for several parameters we fit in {\tt alf}, 
 including the recession velocity, log(age/Gyr) and [Fe/H] \citep{corner}.  
 As described in Section~2, the input spectra have been smoothed to the desired 
 velocity resolutions, which are $110$~km/s for DF~7, $125$~km/s for DF~44 and 
 $135$~km/s for DF~17.  Therefore, when we fit the spectra with {\tt alf}, the 
 velocity dispersion of our model spectra cannot be taken as a description of 
 the intrinsic velocity dispersion of our targets.  To avoid confusion, we 
 describe this property as `resolution' in Figure~\ref{posterior1}, 
 \ref{posterior2}, and \ref{posterior3}.  
 Although we do not attempt to derive reliable velocity dispersion from 
 the smoothed spectra, it is reassuring to see that the resolution 
 we derive is roughly consistent with the sum in quadrature of 
 the measured velocity dispersion from \citep{vanDokkum2016} and 
 the desired resolution.  For all derived parameters, 
 the values of parameters at $16$th, $50$th and $84$th percentiles of 
 posteriors are shown as dashed lines in the 1D histograms and contours 
 in the 2D histograms.  Outliers are shown as dots.  The blue lines in 
 the 2D histograms mark the values of parameters at minimum $\chi^2$. 

 We find that the recession velocities of DF~7, DF~44, and DF~17 
 are $6600^{+40}_{-26}$~km/s, $6402^{+41}_{-39}$~km/s and $8311^{+43}_{-43}$~km/s.  
 The uncertainties are consistent with the mock data tests.  
 For DF~44, our result is consistent with the recession velocity measured 
 by \citet{vanDokkum2016,vanDokkum2017}.  They measured the kinematics of DF~44 using 
 a $33.5$ hours integration spectra observed by the DEIMOS spectrograph 
 on the Keck II telescope, and obtained a recession velocity of 
 $6398^{+6}_{-6}$~km/s for DF~44. The recession velocity of DF~7 is 
 consistent with the measurement by 
 \citet[][cz$=6587\pm33$~km/s]{Kadowaki2017}. 
 The derived recession velocities confirm 
 that all of the three UDGs are members of the Coma cluster. 

\begin{figure}[t]
\vskip 0.25cm
\centering 
\includegraphics[width=9cm]{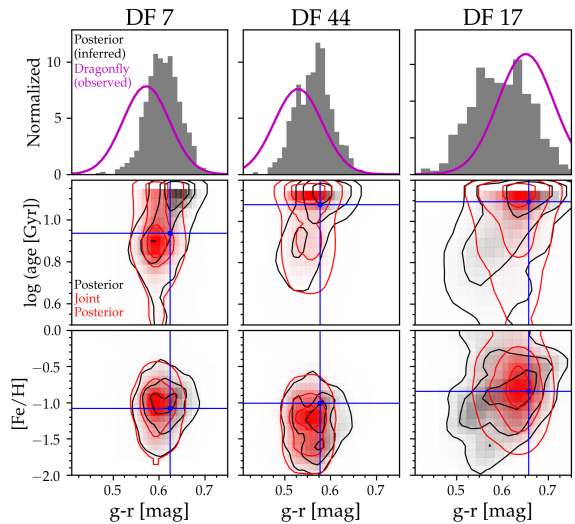}
\caption{
Top panels: 1D histogram of $g-r$ color posterior from {
\tt alf} (gray) for DF~7 (left), DF~44 (middle) and DF~17 (right), 
compared with the $g-r$ color from broadband images 
taken by the Dragonfly Telephoto Array (purple).  Middle and Bottom panels: 
Projections of the posterior of log(age), [Fe/H] and $g-r$ in 2D 
histograms (black). Contours show the 16th, 50th and 84th percentile.  
Blue lines show the best fit parameters at minimum $\chi^2$.  
The joint posterior distributions from combining broadband colors and 
model spectra colors are shown in the same panels in red.
}
\label{colorfig}
\end{figure}
\noindent  

 The ages of DF~7, DF~44 and DF~17 are $8.57^{+4.11}_{-2.93}$~Gyr  
 , $10.50^{+3.00}_{-4.39}$~Gyr and $7.61^{+5.09}_{-4.72}$~Gyr.  
 The iron abundances, [Fe/H], are $-1.03^{+0.31}_{-0.34}$, 
 $-1.25^{+0.33}_{-0.39}$ and $-0.83^{+0.56}_{-0.51}$, respectively.  
 The uncertainties of both log(age/Gyr) and [Fe/H] are similar to what was 
 derived from the mock tests.  From the 2D posterior distributions 
 in Figure~\ref{posterior1}, Figure~\ref{posterior2} and Figure~\ref{posterior3}, 
 one can see that there are no strong degeneracies between the resolution and 
 age or metallicity.  There is a modest degeneracy between age and metallicity, 
 as expected \citep{Worthey1994}.  
 The age posteriors are not perfectly Gaussian, but the local maximum in the 
 marginalized age posteriors are all in the old age regime.  
 The probabilities that the stellar populations 
 are young are very low.  The $16$th percentiles of age posteriors are 
 $5.6$, $6.0$ and $2.9$~Gyr, and the $2.5$th percentiles of age posteriors are 
 $2.6$, $2.7$ and $1.4$~Gyr for DF~7, DF~44 and DF~17, respectively. 
 In addition, the mock test shows that {\tt alf} is able to recover the true ages 
 for young ($3$~Gyr) stellar population. The primary result of this paper is that all 
 three UDGs are old and metal poor. 
 
\begin{figure}[t]
\vskip 0.15cm
\centering 
\includegraphics[width=8.6cm]{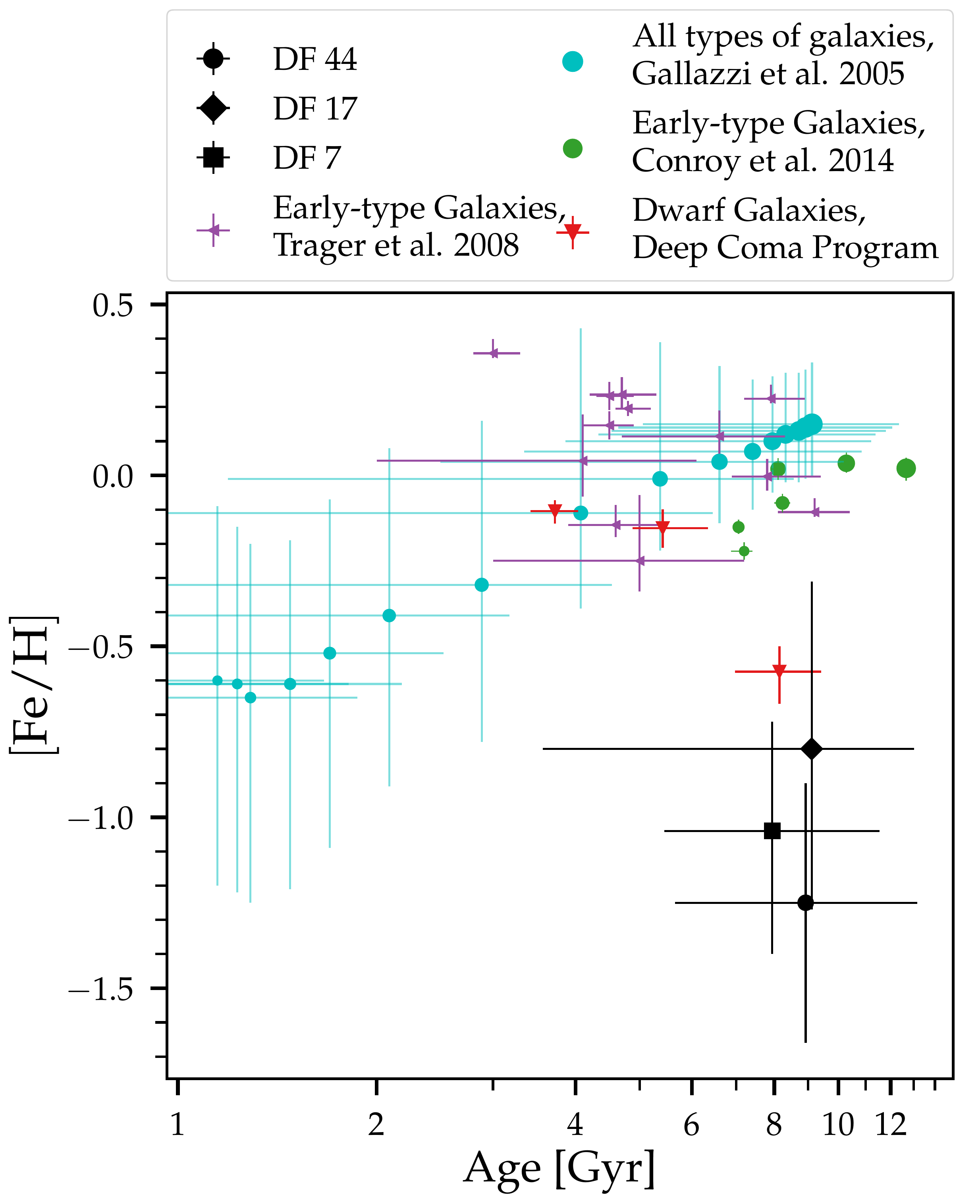}
\caption{
Relation between age and [Fe/H] for DF~44, DF~7 and DF~17 
(black),  three dwarf elliptical galaxies in the Coma cluster 
(red, Gu et al. in prep), and 12 early-type galaxies in the 
Coma cluster measured by \citet{Trager2008}.
In addition, cyan and green data points show the relation for 
various types of galaxies from \citet{Gallazzi2005}, and 
early-type galaxies binned in stellar mass \citep{Conroy2014}.  
For data from \citet{Conroy2014} and \citet{Gallazzi2005}, 
the sizes of data points represent different stellar 
masses of data bins, and they show a general trend that 
more massive galaxies are older and more metal rich. 
}
\label{age-FeH}. 
\end{figure}
\noindent  

\subsection{Combined Constraints from Spectra and Photometry}

 In this section we add the $g-r$ color as an additional constraint to our 
 stellar population parameters.  We measure 
 the color from the Dragonfly data within an aperture of 6\asec, which 
 is the similar to the regions of our stacked spectra.  
 The measured $g-r$ colors are $0.57\pm0.05$~mag, $0.53\pm0.05$~mag and $0.65\pm0.06$~mag, respectively.
 In Figure~\ref{colorfig}, the normalized 1D posterior distributions of the 
 $g-r$ color derived from fitting the continuum-normalized spectra are shown 
 in the top panels in black. 
 We assume the probability density of the observed $g-r$ colors as a normal 
 distribution, and take the measured color and uncertainty as the mean and 
 standard deviation of this normal distribution.  The top panels show that 
 the differences between the observed color and the inferred color 
 from the models are no more than 0.02~mag for both DF~7 and DF~44.  The  
 good agreement between the inference from fitting spectra and the observed 
 colors suggests that reddening is low in both UDGs, and therefore their gas 
 and dust content is likely low. 
 For DF~17, the color from photometry is slightly redder than the color from the 
 spectral models, but the difference is smaller than $1\sigma$ uncertainty.  
 Near-Infrared photometry would be helpful to confirm whether DF~17 is dust reddened. 

 We use the observed $g-r$ color to further constrain the model space.  
 We re-weight the MCMC chains based on the probability density of the 
 broadband $g-r$ color, and generate new posterior distributions by 
 bootstrap resampling.  The middle and bottom panels of 
 Figure~\ref{colorfig} show the joint posterior distributions 
 of log(age/Gyr) and [Fe/H] in red, respectively.  
 Compared with the posterior distributions obtained from spectroscopy 
 alone, we find that the inclusion of photometry results in 
 slightly tighter constraints for DF~44 and DF~7.  This is at least 
 partially due to the relatively large errors on the observed 
 colors - more precise colors would likely result in stronger 
 constraints on the age and metallicity.  The results from these 
 joint constraints are also shown in Table~1.  
 For DF~17, an additional constraint from photometry provides us 
 a slightly older and less metal poor stellar population. 

\subsection{Stellar Mass}
 We calculate the stellar mass using the $g$ band total integrated 
 magnitude from \citet{vanDokkum2014a}. The $g$ band integrated 
 magnitudes for DF~7, DF~44 and DF~17 are $-16.0_{-0.2}^{+0.2}$~mag, 
 $-15.7_{-0.2}^{+0.2}$~mag and $-15.2_{-0.2}^{+0.3}$~mag.  
 The $g-r$ color within 6\asec ~from the galaxy centers are 
 $0.57\pm0.05$~mag, $0.53\pm0.05$~mag and $0.65\pm0.06$~mag.
 We adopt the $r$ band solar absolute magnitude as 4.76 from \citet{Blanton2003}. 
 $r$ band K-corrections are calculated based on the best-fit model spectra, 
 and $K_r\sim -0.02$.  The rest-frame $r$ band mass-to-light ratio for 
 DF~7, DF~44 and DF~17 are taken from the results constrained jointly by 
 spectra and photometry within a radius of 6\asec, and they are 
 $1.56^{+0.47}_{-0.28}$~\solmass/\solum\ , $1.64^{+0.54}_{-0.38}$~\solmass/\solum 
 and $1.80^{+0.51}_{-0.66}$~\solmass/\solum.  Therefore,  
 the stellar masses for the three UDGs are 
 $5.24_{-1.22}^{+2.27} \times 10^8$\solmass~, 
 $4.03_{-1.05}^{+1.87} \times 10^8$\solmass~ 
 and $3.11_{-1.15}^{+1.28} \times 10^8$\solmass, respectively.  
 \citet{vanDokkum2016} has calculated the stellar mass 
 for DF~44 using its $i$-band luminosity and $g-i$ color.
 Our result is consistent with theirs 
 ($M_* \approx 3 \times 10^8$\solmass).

\begin{figure*}[t]
\vskip 0.15cm
\centering 
\includegraphics[width=18cm]{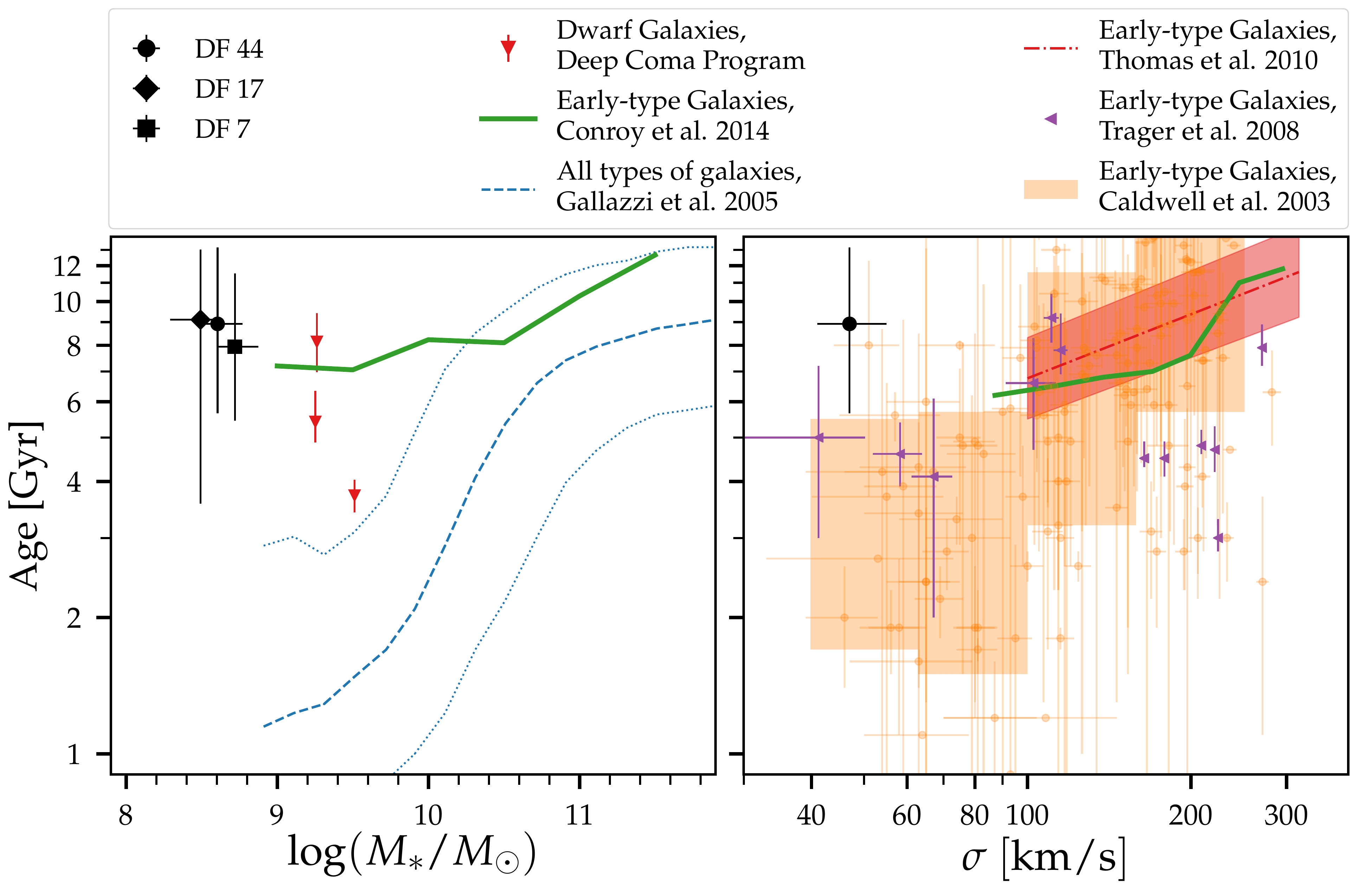}
\caption{
Relationship between age and stellar mass (left panel), 
and velocity dispersion (right panel). Our results are compared to three 
dwarf elliptical galaxies in the Coma cluster (red triangles, Gu et al. 2018, in preparation), early-type galaxies \citep{Conroy2014} binned in stellar mass 
(left panel) and velocity dispersion (right panel), a sample of nearby early-type galaxies 
that mostly with low velocity dispersions \citep{Caldwell2003}, 
morphologically selected SDSS early-type galaxies \citep{Thomas2010}, 
SDSS galaxies covering a wide range of galaxy types \citep{Gallazzi2005}, 
and 12 early-type galaxies in the Coma cluster \citep{Trager2008}.
}
\label{UDGage}
\end{figure*}
\noindent  
\begin{figure*}[t]
\vskip 0.15cm
\centering 
\includegraphics[width=18cm]{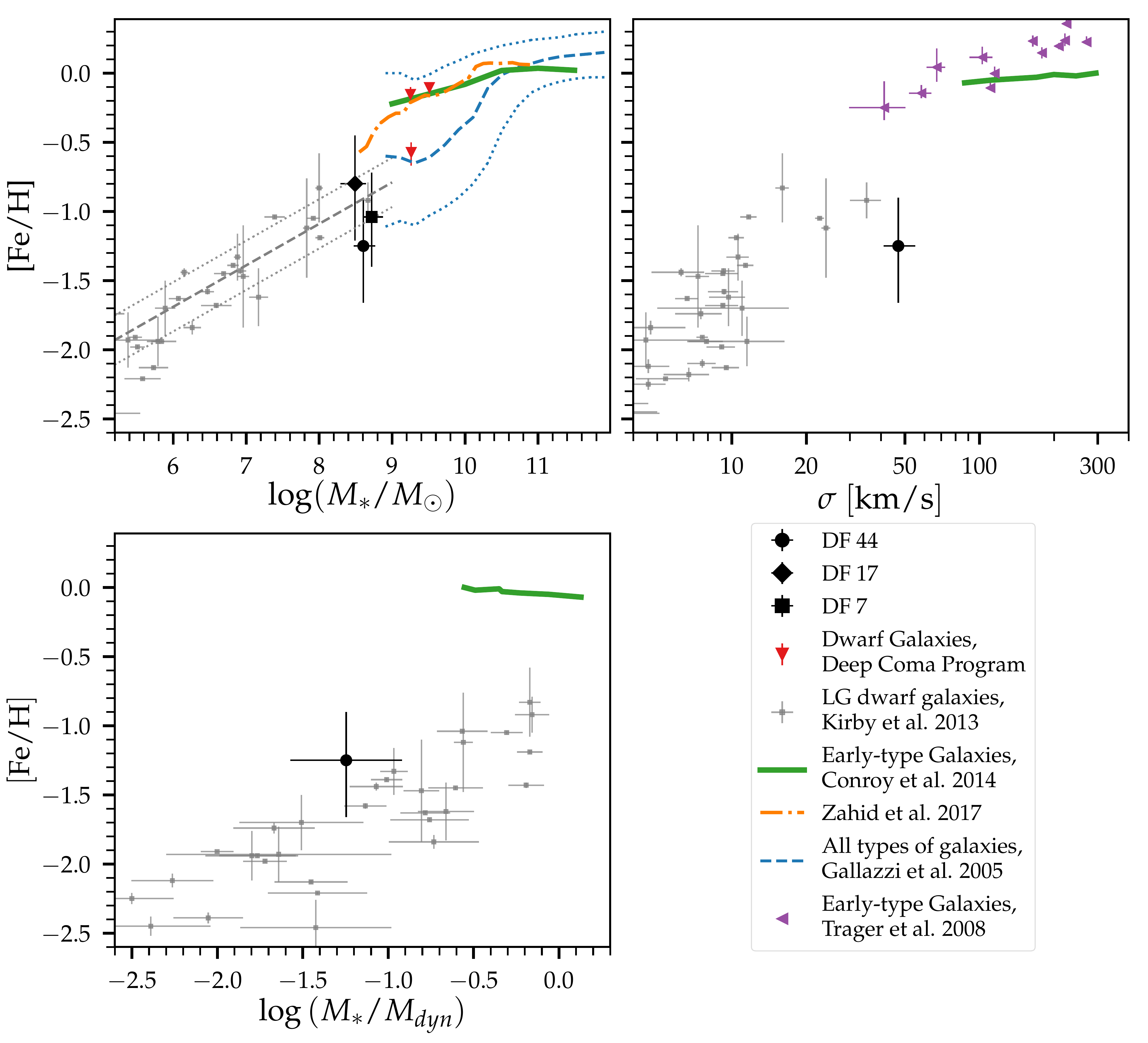}
\caption{
Top left: relation between stellar mass and [Fe/H] for DF~44, DF~7, and DF~17
(black), compared with the stellar mass$–-$stellar metallicity 
relation for three dwarf elliptical galaxies in the Coma cluster 
(triangles, Gu et al. 2018, in preparation), 
and previous results from the literature.  Gray symbols show Local Group dwarf 
galaxies from \citet{Kirby2013}, and gray dashed and dotted lines represent the 
median and 16th and 84th percentiles of the metallicity distributions.
Cyan dashed and dotted lines show the median and 16th and 84th 
percentiles of the metallicity distributions for various types of 
galaxies in \citet{Gallazzi2005}. Green symbols show the stellar 
mass$–-$metallicity relation for early-type galaxies binned in stellar mass \citep{Conroy2014};
~~Top right: relation between velocity dispersion and [Fe/H] for DF~44, 
compared with early-type galaxies binned in velocity dispersion \citep{Conroy2014}, 
and Local Group dwarf galaxies \citep{Kirby2013};  
~~Bottom left: Relation between metallicity and the logarithmic ratio of 
stellar mass to dynamical mass for DF~44, Local Group dwarf galaxies 
\citep{Kirby2013} and early-type galaxies binned in velocity 
dispersion \citep{Conroy2014}.
}
\label{Kirby2013}. 
\end{figure*}
\noindent  

\section{Discussion}
 
 We have presented the first spectroscopic measurements of age and metallicity 
 for three UDGs in the Coma cluster, DF~7, DF~44 and DF~17. In addition, we have 
 presented the first spectroscopic redshift for DF~17.  These three UDGs are among 
 the brightest and largest UDGs in the sample of \citet{vanDokkum2015}.  We find 
 that all of the three UDGs 
 are old ($7.94^{+3.60}_{-2.49}$~Gyr, $8.92^{+4.25}_{-3.26}$~Gyr and 
  $9.11^{+3.91}_{-5.54}$~Gyr), and metal poor 
  ($\mathrm{[Fe/H]_{DF~7}}=-1.04^{+0.32}_{-0.36}$, 
   $\mathrm{[Fe/H]_{DF~44}}=-1.25^{+0.35}_{-0.41}$,
   $\mathrm{[Fe/H]_{DF~17}}=-0.80^{+0.49}_{-0.41}$).  
 Their stellar masses are $5.24_{-1.22}^{+2.27} \times 10^8$\solmass~, 
 $4.03_{-1.05}^{+1.87} \times 10^8$\solmass~ 
 and $3.11_{-1.15}^{+1.28} \times 10^8$\solmass, 
 respectively. 
 The results are summarized in Figure~\ref{age-FeH}.
 The derived ages and metallicities are typical for 
 low mass galaxies in cluster environments 
 \citep[e.g.,][]{vanZee2004, Penny2007, Smith2009}.  
 \citet{Kadowaki2017} concluded that on average four Coma cluster 
 UDGs in their sample are metal poor by visually comparing the stacked 
 spectrum against SSP models.  The UDGs in our samples have metallicities 
 consistent with this conclusion.
 Their ages are slightly older than early-type galaxies with similar 
 or larger stellar mass.  Their metallicities are broadly consistent 
 with dwarf galaxies at similar stellar masses.

 The ages of DF~7, DF~44 and DF~17 show that they all have old stellar 
 populations, indicating that the star formation in all of the three 
 galaxies must have been truncated at high redshift. 
 In Figure~\ref{UDGage}, we show relationships between age, stellar mass, and
 velocity dispersion of the UDGs, and compare to other galaxies from the 
 literature.  The left panel shows three dwarf ellipticals in the Coma cluster 
 obtained as part of the Deep Coma program  
 (Gu et al. in prep).  Their stellar population properties are also derived 
 from full optical spectral modeling with {\tt alf}. We also plot 
 stellar properties of early-type galaxies from stacked SDSS spectra by 
 \citet{Conroy2014}.   Amongst early-type galaxies there is a weak trend such 
 that less massive galaxies are slightly younger.  In addition, we compare with a  
 large magnitude-limited sample of SDSS galaxies from \citet{Gallazzi2005}.  
 Note that in their sample the low mass galaxies are dominated by late-type 
 galaxies and hence the mean ages are much younger.
 
 In the right panel of Figure~\ref{UDGage}, we plot the relation between age 
 and velocity dispersion. \citet{Thomas2010} studied a large sample of 
 morphologically selected early-type SDSS galaxies. 
 \citet{Caldwell2003} studied a sample including mostly low velocity 
 dispersion ($\sigma<100$~km$^{-1}$) early-type galaxies in the Virgo Cluster 
 and in lower density environments.  They found that there is 
 a general trend that the galaxies with higher velocity dispersion are older.  
 \citet{Trager2008} studied 12 early-type galaxies in the Coma cluster, 
 including both elliptical and S0 galaxies. They seem to broadly obey the 
 same (weak) relation.  Both panels seem to suggest that, despite different 
 techniques for estimating ages, there is a general trend that galaxies 
 with higher stellar masses or larger velocity dispersions are older. 
 Although the uncertainties are large, the three UDGs in this paper do 
 not seem to fall on the same age--stellar mass trend.  We only have velocity 
 dispersion for DF~44 \citep[][$\sigma=47^{+8}_{-6}$~km$^{-1}$]{vanDokkum2017}.  
 DF~44 also seems to be an outlier on the apparent age--$\sigma$ trend.
 They are apparently older than dwarf elliptical and S0 galaxies
 at similar stellar mass and velocity dispersion in the Coma cluster 
 \citep{Trager2008}.  
 
 The old ages indicate that the stellar components of large and diffuse 
 galaxies such as DF~7, DF~44, and DF~17 can be formed at high redshift.  
 Although we have only investigated
 three UDGs that are all at the high end of the UDG size distribution, our 
 results seem to rule out the scenario where UDGs recently quenched their 
 star formation, for instance, due to recent accretion onto the Coma cluster. 
 These results are also not consistent with the theoretical 
 prediction that UDGs have a relatively late formation time compared to other 
 low-mass galaxies \citep[e.g.,][]{Rong2017}.  It is worth noting that 
 our single-age assumption in the stellar population modeling is not 
 enough to disentangle the full star formation 
 history; future measurements of properties such as $\alpha$-abundance 
 will provide us with more information about the formation of UDGs. 
 
 Recent studies have shown that at least some of the UDGs have rich GC systems. \citet{Peng2016} estimated that the number of GCs in DF~17 is $28\pm14$, and \citet{vanDokkum2017} estimated 
 that DF~44 hosts $74\pm18$ GCs.  The specific frequency is 
 $S_N=26\pm13$ for DF~17 \citep{Peng2016}, and $S_N=27\pm7$ for DF~44 
 \citep{vanDokkum2017}.  Since GCs are formed during early and rapid star 
 formation, the relative higher specific frequency of GCs in these two 
 UDGs may reflect that they have experienced intense starburst at high 
 redshift \citep[e.g.,][]{Liu2016}.  
 The old ages derived for DF~44 and DF~17 are consistent with this picture.   
 
 The relation between age and metallicity is shown in Figure~\ref{age-FeH}.  
 The three UDGs are compared with three dwarf elliptical 
 galaxies in the Coma cluster, 12 early-type galaxies in the Coma cluster from 
 \citet{Trager2008}, massive early-type galaxies binned in stellar mass from 
 \citet{Conroy2014}, and SDSS galaxies covering a wide range of galaxy types 
 from \citet{Gallazzi2005}.  For data from \citet{Conroy2014} and 
 \citet{Gallazzi2005}, the sizes of data points represent different stellar 
 masses of data bins, and they show a general trend that more massive 
 galaxies are older and more metal rich. The UDGs appear as outliers.  
 Their low metallicities and old ages are consistent with a picture that 
 their star formation histories are brief at high redshifts.  Three dwarf 
 elliptical galaxies in the Coma cluster appear to lie between UDGs and 
 more massive galaxies.  It would be interesting to look into the star 
 formation and chemical enrichment histories of a larger sample of dwarf 
 galaxies and UDGs in order to investigate whether there is any connection 
 between these two properties.
 
 The stellar mass--stellar metallicity relation (MZR) for galaxies 
 provides important clues to their star formation and chemical enrichment 
 history. The fact that this relation has relatively low scatter is a challenge 
 to explain, especially at the low-mass end \citep[e.g.,][]{Kirby2013}.  
 Previous work suggests that this relation is linked to the complex 
 interplay between reionization, star formation, gas inflow, outflow, and recycling 
 \citep[e.g.,][]{Ma2015,Lu2017}.  

 The top-left panel of Figure~\ref{Kirby2013} shows the locations of DF~7, 
 DF~44 and DF~17 on the MZR relative to other populations, including dwarf galaxies 
 in the Local Group \citep{Kirby2013} (gray dots), early-type galaxies stacked 
 in stellar mass bins \citep{Conroy2014} (green), star forming galaxies in SDSS 
 \citep{Zahid2017}, and the MZR from 
 \citet{Gallazzi2005} which covers both star forming and quiescent SDSS 
 galaxies (light blue).  
 We note that the data points from \citet{Zahid2017} and \citet{Gallazzi2005} 
 in Figure~\ref{Kirby2013} represent the total metallicities instead of 
 iron abundances of galaxies in their samples.
 The large scatter of the MZR from \citet{Gallazzi2005} 
 is at least partially due to the mixture of both early- and late-type galaxies 
 at lower masses combined with the increased difficulty of measuring stellar 
 metallicities for star-forming galaxies.  Considering that effect, the MZR seems to be 
 continuous from low to high masses.  Triangle symbols 
 show three dwarf elliptical galaxies from our sample in the Coma cluster 
 (Gu et al. 2018, in preparation), whose metallicities are generally consistent with 
 the MZR from \citet{Kirby2013}, \citet{Conroy2014}, and \citet{Gallazzi2005}.  
 Despite the large uncertainties of metallicity and stellar mass, all of 
 the three UDGs seem to follow the MZR defined by normal dwarf galaxies. 
 This suggests that stellar mass plays an important role in determining  
 stellar metallicities, regardless of a galaxy's size. 
 
 Previous work concluded that some UDGs, including 
 DF~44, may live in more massive dark matter halos than would be 
 expected for their luminosities
 \citep[e.g., $M_{\rm vir}>10^{11} M_\odot$,][]{vanDokkum2016}. 
 In addition, it is well-known that the stellar properties of early-type 
 galaxies are tightly related to their dynamical masses \citep[e.g.,][]{Gallazzi2006, Graves2009}.
 Therefore, we also look into the relation between their stellar properties and 
 gravitational potential well depth.  
 The top-right panel of Figure~\ref{Kirby2013} shows the relation between 
 stellar velocity dispersion ($\sigma$) and metallicities.  We compare 
 DF~44 with dwarf galaxies \citep{Kirby2013}, massive early-type galaxies 
 binned in velocity dispersion \citep{Conroy2014}.  We also compare to  
 12 elliptical and S0 galaxies in the Coma cluster \citep{Trager2008}.  
 Their iron abundances are calculated based on their metallicities and 
 the enhanced element abundances ([Z/H] and [E/Fe]) in \citet{Trager2008} 
 and Eq~3 in \citet{Trager2000}.
 DF~44 stands out in this plot in the sense that it has lower metallicity 
 for its $\sigma$ compared to other galaxies.  We also know that the stellar 
 mass of DF~44 is unusually low for its $\sigma$ and since it is the 
 (massive) stars that produce metals, it is perhaps not surprising that 
 DF 44 has low metallicity for its $\sigma$. 
 
 We explore this point further in the bottom-left panel of 
 Figure~\ref{Kirby2013}, where we plot [Fe/H] as a function of the logarithmic 
 ratio of stellar mass to dynamical mass.  This ratio is proportional to the 
 integrated star formation efficiency.  In this diagram, one can imagine 
 diagonal tracks where a galaxy in a given halo evolves along the track as 
 it converts baryons into stars.  Tracks associated with more massive halos 
 will be offset vertically in this diagram, as a given integrated star 
 formation efficiency (a given stellar-to-dynamical mass ratio) will result 
 in a higher stellar metallicity for galaxies in deeper potential wells, as 
 such systems will be able to retain a greater fraction of the metals.
 
 In this diagram, we compare the location of DF~44 to dwarf galaxies 
 in the Local Group \citep{Kirby2013} and massive early-type galaxies 
 binned in velocity dispersion \citep{Conroy2014}. 
 The dynamical masses of all galaxies in this panel are estimated 
 using Eq~2 in \citet{Wolf2010} for dispersion-supported systems:
\begin{equation}
M_{dyn} = 4\sigma^2R_e/G\,.
\end{equation}
As anticipated, for a given stellar-to-dynamical mass ratio, galaxies in 
more massive halos have higher metallicities.    
DF~44 lies slightly above Local Group dwarf galaxies in this plot.  
Although the offset is marginal, 
it could be consistent with the idea that DF 44 is in an unusually massive 
dark matter halo for its stellar mass, which would allow this galaxy to 
retain a greater fraction of metals ejected from massive stars and hence 
result in a higher stellar metallicity.  Further observations, including 
higher S/N spectra, will be necessary to make stronger conclusions.
 
 Although DF~7, DF~44 and DF~17 reside in the inner regions of the Coma 
 cluster, Figure~\ref{Kirby2013} is not consistent with the picture that 
 UDGs are tidal debris of more massive systems, unless the massive progenitors 
 are true outliers of the stellar mass--stellar metallicity relation with high 
 stellar masses and very metal poor stellar populations.  
 Also, their regular and smooth morphology do not show evidence of tidal disruption. 
 
 In this work, we have focused on three relatively luminous and large UDGs 
 in the cluster environment.  More data are needed on a wider variety of 
 systems in order to make more general conclusions regarding the formation 
 pathway(s) of UDGs.  

\section{Summary}

 We have presented the first stellar population analysis for three 
 UDGs in the Coma cluster, DF~7, DF~44, and DF~17 based on the analysis of 
 their optical spectra.  We have measured their recession velocities, 
 ages, metallicities, and stellar masses using spectra obtained as part 
 of the Deep Coma Program within the SDSS-IV/MaNGA survey.  We confirm 
 that all of the three UDGs are members of the Coma cluster.  
 They are all old and metal poor, with ages of  
 $7.94^{+3.60}_{-2.49}$~Gyr, $8.92^{+4.25}_{-3.26}$~Gyr, and 
  $9.11^{+3.91}_{-5.54}$~Gyr, and 
  iron abundance of $\mathrm{[Fe/H]}=-1.04^{+0.32}_{-0.36}$, 
   $\mathrm{[Fe/H]}=-1.25^{+0.35}_{-0.41}$,
   $\mathrm{[Fe/H]}=-0.80^{+0.49}_{-0.47}$, respectively.  
  Their stellar masses are $5.24_{-1.22}^{+2.27} \times 10^8$\solmass~, 
  $4.03_{-1.05}^{+1.87} \times 10^8$\solmass~, 
  and $3.11_{-1.15}^{+1.28} \times 10^8$\solmass.
 Their stellar populations are slightly older than early-type galaxies 
 with similar or larger stellar mass and metallicity.  Their metallicities 
 are broadly consistent with known dwarf galaxies at similar stellar masses, 
 but DF~44 falls below the $\mathrm{[Fe/H]}$--$\sigma$ relation.  
  In spite of their surprisingly diffuse structures and large sizes, 
  it appears that their basic stellar population properties are not very 
  atypical for their masses.  These results disfavor UDG formation scenarios 
  that predict late star formation and/or late quenching.   
   
\acknowledgments
We thank the referee for helpful comments that improved this paper.
M.G. acknowledges support from the National Science Foundation 
Graduate Research Fellowship.  
C.C. acknowledges support from NASA grant NNX15AK14G, NSF grant
AST-1313280, and the Packard Foundation. 
M.B. acknowledges funding from NSF/AST-1517006.
A.W. acknowledges support of a Leverhulme Trust Early Career Fellowship.
The computations in this paper were run on the Odyssey cluster supported 
by the FAS Division of Science, Research Computing Group at Harvard University.

Funding for the Sloan Digital Sky Survey IV has been provided by the 
Alfred P. Sloan Foundation, the U.S. Department of Energy Office of
Science, and the Participating Institutions. SDSS-IV acknowledges 
support and resources from the Center for High-Performance Computing 
at the University of Utah. The SDSS website is www.sdss.org.

SDSS-IV is managed by the Astrophysical Research Consortium for the 
Participating Institutions of the SDSS Collaboration including the 
Brazilian Participation Group, the Carnegie Institution for Science, 
Carnegie Mellon University, the Chilean Participation Group, the 
French Participation Group, Harvard-Smithsonian Center for Astrophysics, 
Instituto de Astrof\'isica de Canarias, The Johns Hopkins University, 
Kavli Institute for the Physics and Mathematics of the Universe 
(IPMU) / University of Tokyo, Lawrence Berkeley National Laboratory, 
Leibniz Institut f\"ur Astrophysik Potsdam (AIP),  
Max-Planck-Institut f\"ur Astronomie (MPIA Heidelberg), 
Max-Planck-Institut f\"ur Astrophysik (MPA Garching), 
Max-Planck-Institut f\"ur Extraterrestrische Physik (MPE), 
National Astronomical Observatories of China, New Mexico State University, 
New York University, University of Notre Dame, 
Observat\'ario Nacional / MCTI, The Ohio State University, 
Pennsylvania State University, Shanghai Astronomical Observatory, 
United Kingdom Participation Group, Universidad Nacional 
Aut\'onoma de M\'exico, University of Arizona, 
University of Colorado Boulder, University of Oxford, 
University of Portsmouth, University of Utah, University of Virginia, 
University of Washington, University of Wisconsin, 
Vanderbilt University, and Yale University.

\end{document}